\documentclass{jfm}

\usepackage{graphicx}
\usepackage{epstopdf, epsfig}

\usepackage{subfig}
\usepackage{floatrow}

\graphicspath{{picture/}}

\shorttitle{Nonlinear optimal control of bypass transition}
\shortauthor{D. Xiao and G. Papadakis}

\title{Nonlinear optimal control of bypass transition using a receding horizon approach}

\author{Dandan Xiao,
 George Papadakis
 \corresp{\email{g.papadakis@ic.ac.uk}}}

\affiliation{Department of Aeronautics, Imperial College, London SW7 2AZ, UK}

\begin{document}

\floatsetup[figure]{style=plain,subcapbesideposition=top}

\maketitle

\begin{abstract}
This paper considers the nonlinear optimal control of bypass transition in a boundary layer flow subjected to \textcolor{black}{ a pair of} free stream vortical perturbations using a receding horizon approach. The optimal control problem is solved using the Lagrange variational technique that results in a set of linearised adjoint equations, which are used to obtain the optimal wall actuation (blowing and suction from a control slot located in the transition region). The receding horizon approach enables the application of control action over a longer time period, and this allows the extraction of time-averaged statistics as well as investigation of the control effect downstream of the control slot. The results show that the controlled flow energy is initially reduced in the streamwise direction and then increased because transition still occurs. The distribution of the optimal control velocity responds to the flow activity above and upstream of the control slot. The control effect propagates downstream of the slot and the flow energy is reduced up to the exit of the computational domain. The mean drag reduction is $55\%$ and $10\%$ in the control region and downstream of the slot, respectively. The control mechanism is investigated by examining the second order statistics and the two-point correlations. It is found that in the upstream (left) side of the slot, the controller counteracts the near wall high speed streaks and reduces the turbulent shear stress; this is akin to opposition control in channel flow, and since the time-average control velocity is positive, it is more similar to blowing-only opposition control. In the downstream (right) side of the slot, the controller reacts to the impingement of turbulent spots that have been produced upstream and inside the boundary layer (top-bottom mechanism). The control velocity is positive and increases in the streamwise direction, and the flow behavior is similar to that of uniform blowing. 

\end{abstract}


\section{Introduction}\label{sec:introduction}
In boundary layers, transition from laminar to turbulent flow is usually classified as orderly (classic) or bypass. Orderly transition is a slow process, which involves the exponential growth of Tollmien-Schlichting (TS) waves, their secondary instability and finally breakdown to turbulence. On the other hand, bypass transition refers to all other routes. In this work, we focus on bypass transition in a zero-pressure-gradient flat-plate boundary layer flow exposed to \textcolor{black}{a pair of} vortical perturbations.

Laminar-turbulent transition in boundary layer flow is always associated with high skin friction and enhanced mixing of momentum. Most of the turbulent energy generation and dissipation takes place inside the boundary layer and thus affects engineering performance. In particular, bypass transition is found in gas turbines, where high levels of free stream turbulence are present \citep{hodson2005bladerow, zaki2010direct,wissink2014effect}. Therefore exploring possible control methods to delay or suppress bypass transition is of crucial importance for performance improvement.

\subsection{Active control of wall-bounded flows}
Active control methods provide better control performance and have been explored widely in the literature. One simple type of active control is uniform blowing (UB) or suction (US) at the wall. \cite{park1999effects} performed DNS with steady UB or US with velocity magnitude less than $10\%$ of the mean flow velocity from a narrow spanwise slot in a turbulent boundary layer flow. They found that for UB, skin friction over the control slot decreased rapidly but increased downstream due to enhanced tilting and stretching of vorticity. Above the slot, streamwise vortices are lifted up and weakened by the actuation. The opposite is found for US. \cite{kametani2011direct} examined the effect of UB and US with much smaller velocity magnitudes (less than $1\%$ of the mean flow velocity) in a spatially developing turbulent boundary layer using DNS. The control slot covered the entire computational domain. They applied the FIK identity \citep{fukagata2002contribution} to investigate the drag reduction mechanism and concluded that the mean convection has a strong contribution in reducing the drag for UB and increasing the drag for US. More recently, \cite{kametani2015effect} applied UB and US in a turbulent boundary layer at moderate Reynolds number using large eddy simulations (LES). The actuation velocity had a magnitude of $0.1\%U_\infty$ and achieved more than $10\%$ drag reduction (or enhancement) by UB (or US). They also found that the drag reduction efficiency could be improved by using a wider control region which starts at a more upstream location.

Many control strategies on drag reduction are guided by the underlying flow physics. For example, \cite{choi1994active} proposed the opposition control method to suppress the coherent structures in the wall region in a turbulent channel flow. They imposed a transpiration velocity which is equal and opposite to the wall-normal velocity at a detection plane located at some distance from the wall, in order to counteract the motion of streamwise vortices. They found that drag is initially reduced as a result of suppressed sweep events, and at later times drag reduction was due to the change of wall vorticity layer by the active control. \cite{stroh2015comparison} compared the performance of opposition control in a turbulent channel flow and a spatially developing turbulent boundary layer. They found that for both configurations the drag reduction rates were similar, but the mechanism was different. In channel flow, drag is reduced due to the attenuation of the Reynolds stress, while in a boundary layer modification of the spatial flow development is of critical importance. 

\cite{chang2002viscous} investigated the effect of Reynolds number on opposition control using LES and showed that drag reduction is less effective at higher Reynolds number. \cite{chung2011effectiveness} studied the effect of the amplitude of opposition control as well as the location of the detection plane on the control performance and found that drag reduction is proportional to the magnitude of the blowing and suction velocity.

Alternative strategies to the full opposition control have also been developed. \cite{pamies2007response} modified the opposition control, whereby only the blowing part was retained and the suction part was suppressed. They applied this blowing-only opposition control using LES in a spatially developing turbulent boundary layer and demonstrated that drag reduction efficiency is improved. Recently, \cite{abbassi2017skin} performed control experiments in a high Reynolds number turbulent boundary layer using also the opposition control framework. They used a spanwise array of wall-normal jets, which were activated based on information of incoming high-speed flow zones. The latter were detected with wall shear stress sensors placed upstream of the actuators. The wall-normal jet velocity opposed only the down-wash action of the natural, large scale roll-modes. \cite{lee1997application} used a neutral network to predict the opposition blowing and suction actuations based on wall shear stress and achieved $20\%$ skin friction reduction in a channel flow. They observed a stable pattern in the distribution of the weights from the neutral network and derived a simple control scheme based on a local weighted sum of spanwise shear stress.

There are several other successful active control methods such as spanwise wall oscillations \citep{quadrio2004critical,yudhistira2011direct,lardeau2013streamwise,hack2014influence} and wall-deformation \citep{nakanishi2012relaminarization,tomiyama2013direct}, but they are outside the scope of the present work.

\subsection{Nonlinear optimal control}
Most recent control approaches in delaying transition inside boundary layers are based on linear control models (\cite{chevalier2007linear,monokrousos2008dns,papadakis2016closed,bagheri2009input}). The streamwise streaks, generated inside the boundary layer by the free stream vortical disturbances, initially grow linearly. In later stages however, when they breakdown to form turbulent spots, the nonlinear effects become important. Very few nonlinear approaches for controlling transition in a boundary layer have appeared in literature. These are gradient-based approaches i.e.\ the control parameters are updated along the direction that minimizes a given cost function. The gradient direction is obtained by solving the governing equations (i.e.\ Navier-Stokes equations) and the adjoint equations in a forward-backward iterative loop. Since the optimal control inputs are solely based on the governing equations and the objective function, they do not rely on physical intuition \citep{flinois2015optimal}. Instead, the optimal control solution, which gives the best achievable performance, can be used a posteriori to derive an effective control strategy. 

The aforementioned approach is also known as model predictive control (MPC). \citet{cherubini2013nonlinear} applied MPC to a three-dimensional boundary layer flow with optimally growing perturbations as initial conditions and successfully brought the flow back to the laminar state. They also compared the performance of linear and nonlinear controllers. The results showed that the fully nonlinear controller was more effective than the linear one in reducing the energy of the finite-amplitude perturbations. \citet{xiao2017nonlinear} applied MPC for a short optimization horizon to suppress bypass transition in a boundary layer; the controller was effective in reducing the flow energy above the control slot. Applications of optimal control based on adjoint methods to other flow problems can also be found in the literature. For instance, \cite{mao2015nonlinear} applied this framework to suppress vortex shedding from a circular cylinder using wall normal transpiration.

When MPC is applied to chaotic flows, the optimisation length is limited by the growth of instability of the linear adjoint equations when integrating backwards in time \citep{wang2013forward}. This instability is physical and is due to (one or more) positive Lyapunov exponents that arise when linearising along a solution trajectory.  In order to perform control simulations for longer time, the optimal control approach can be applied in a receding horizon framework. The flow is locally optimised over a finite time period $T$ using the forward/backward iteration loop described above. Once the convergence is achieved within $T$, the flow is advanced in time by some portion $T_a$ of $T$ and a new optimisation problem is solved. The application of this framework to turbulence in channel flow was first introduced by \citet{bewley2001dns}. The authors achieved relaminarization at $Re_{\tau}=180$ (based on mean friction velocity and the channel half-height). \textcolor{black}{\cite{passaggia2013adjoint} applied the adjoint based optimisation procedure to control the dynamics of a separated boundary layer flow over a bump in a two-dimensional setting.} \cite{flinois2015optimal} also designed an adjoint-based optimal control using receding control to suppress vortex shedding in the wake of a circular cylinder via rotation. They examined the control performance for different $T$ and found that long $T$ was essential and resulted in better performance. 

In this study, we apply nonlinear optimal control in a receding horizon framework to suppress bypass transition due to \textcolor{black}{a pair of} free stream vortical disturbances. The paper is organised as follows. In section \ref{sec:methodology}, we introduce the numerical methodology and the nonlinear optimal control algorithm. In section \ref{sec:results}, results from optimal control of bypass transition are presented. In particular, the control results are first investigated through the flow energy and the distribution of the optimal control velocity. Then the control performance is assessed by examining the time-averaged statistics over the control region as well as downstream of the control slot. Two-point correlations between the control velocity and the flow above are studied in order to understand the control mechanism. The main findings are summarised in section \ref{sec:conclusions}. 

\section{Methodology}\label{sec:methodology}
\subsection{Direct numerical simulations}
The flow is governed by the continuity and Navier-Stokes equations. For a three-dimensional incompressible flow, the non-dimensional form of this set equations reads,
\begin{equation}\label{eq: continuity equation}
 \nabla \cdot \textbf{u} =  0,
 \end{equation}
 \begin{equation}\label{eq: Navier-stokes equation}
\frac { \partial \textbf{u}}{ \partial t} = -(\textbf{u} \cdot \nabla)\textbf{u}-\nabla p+\frac{1}{Re_{L_0}} \nabla ^{2}\textbf{u}
\end{equation}

The spatial variables are defined in Cartesian coordinates and non-dimensionalized by the Blasius similarity variable $L_0=\sqrt{\nu x_0^*/ U_{\infty}}$, where $x_0^*$ is the dimensional streamwise distance between the leading edge of the flat plate and the inlet location of the computational domain, $\nu$ is the kinematic viscosity of the fluid and $U_{\infty}$ is the free stream velocity. The velocity vector is non-dimensionalized as $\textbf{u}=\textbf{u}^*/U_{\infty}$, pressure as $p=p^*/\left( \rho U_{\infty}^2 \right)$, where $\rho$ is the fluid density, and time as $t=t^*/\left(L_0/U_{\infty}\right)$.

\begin{figure}
\centerline{\includegraphics[width=0.99\textwidth]{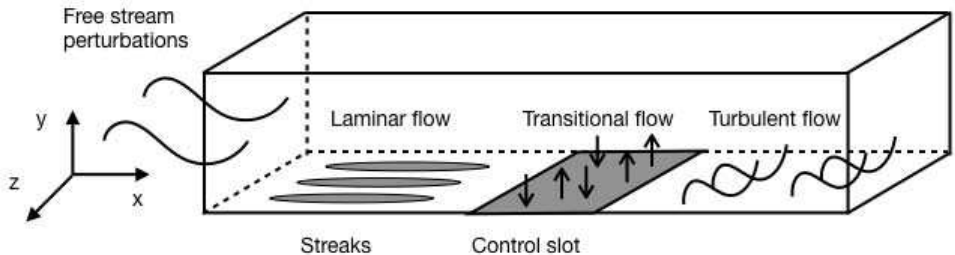}}
\caption{Schematic representation of the computational domain used for the simulation of controlled boundary layer flow over a flat plate subjected to free stream perturbations. Actuation is through blowing and suction from a control slot located in the transition region.}
\label{fig:geometry}
\end{figure}

Figure \ref{fig:geometry} shows the computational domain, which is a rectangular box of dimensions $1800 \times 100 \times 90$. \textcolor{black}{The boundary layer thickness $\delta$ is defined as the distance from the wall at which the velocity reaches 0.99$U_{\infty}$.} The Reynolds number at the inlet is $Re_{L_0}=200$ (based on the momentum thickness $Re_{\theta_0^*}=134$). The grid is uniformly distributed in both streamwise and spanwise directions, but is stretched in the wall-normal direction with an expansion ratio $1.031$. The mesh resolution is $1800 \times 150 \times 120$. 

Bypass transition is triggered by imposing at the inlet boundary vortical perturbations, which are constructed using the continuous modes of the Orr-Sommerfeld and Squire equations \citep{jacobs2001simulations,brandt2004transition}. In the current work, we only use two modes, one of low and one of high frequency to trigger the transition \citep{zaki2005mode}. As will be seen later, this makes the analysis of the control mechanism easier. Parameters for the two modes are listed in table \ref{tab: parameters for two modes}. A detail description of the boundary conditions, numerical code, simulation parameters and validation can be found in \citet{xiao2017nonlinear}. 

\begin{table}
  \begin{center}
\def~{\hphantom{0}}
  \begin{tabular}{cccccc}
       & $\omega$   &   $\alpha$ & $\beta$ & $\gamma$ \\[3pt]
mode A & $0.0064$ & $0.0064+0.0015i$ & $0.393$ & $0.374$\\
mode B & $0.128$ & $0.128+0.0023i$ & $0.628$ & $0.209$\\
  \end{tabular}
  \caption{Parameters of the continuous modes. $\omega$, frequency; $\alpha$, streamwise wavenumber; $\beta$, spanwise wavenumber; $\gamma$, wall-normal wavenumber. }
  \label{tab: parameters for two modes}
  \end{center}
\end{table}

\begin{figure}\centering
\sidesubfloat[]{
\includegraphics[trim={0cm -0.cm 0cm -0.cm}, clip=true,width=0.95\textwidth]{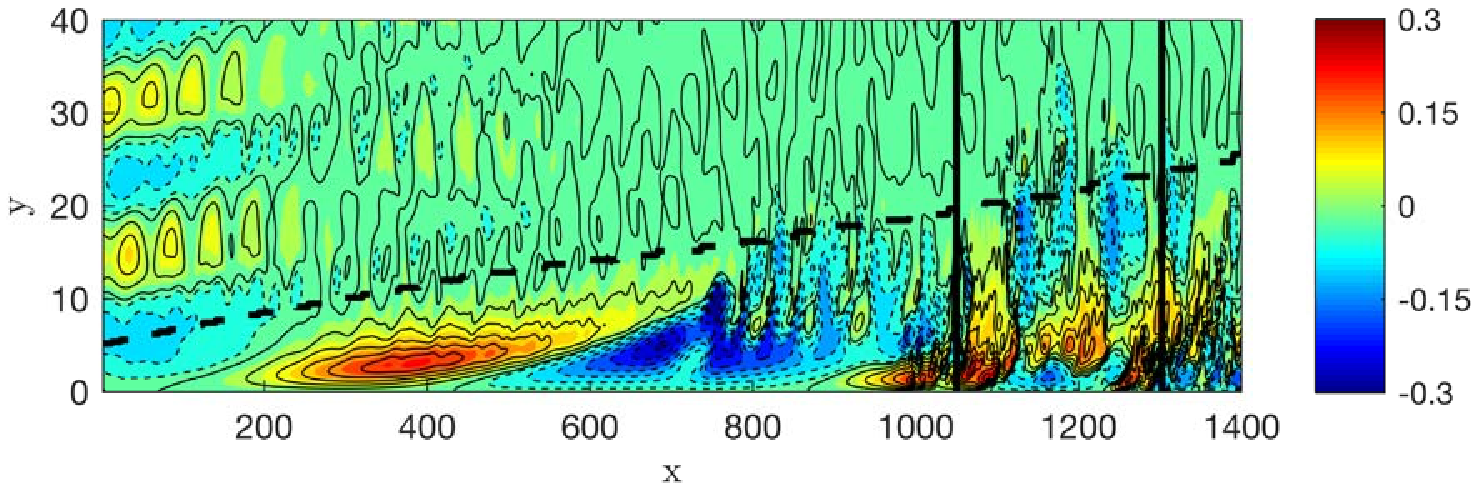}}\quad
\sidesubfloat[]{
\includegraphics[trim={0cm -0.cm 0cm -0.cm}, clip=true,width=0.95\textwidth]{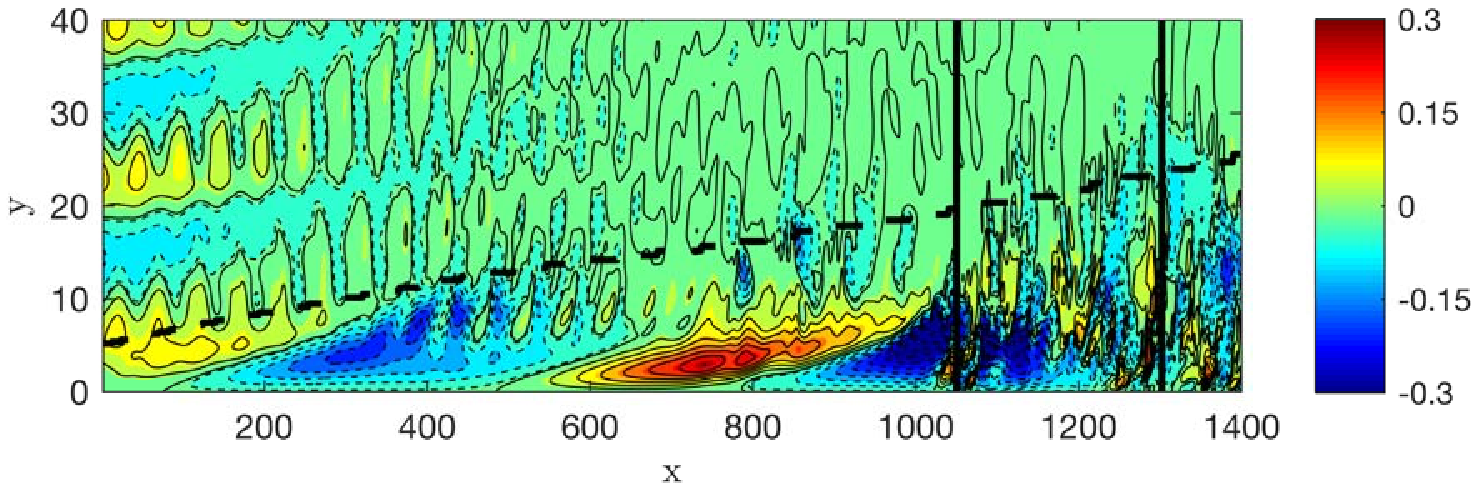}}\quad
\caption{Instantaneous contour plots of streamwise velocity fluctuation in x-y plane at $z=45$ (uncontrolled flow) two time instants (a) 18T; (b) 26T, where $T=60$ (time separation $\Delta t=480$). The black dashed line shows the boundary layer thickness. The vertical black lines mark the location of the control slot, from $1050$ to $1300$ (inactive in this figure).}
\label{fig: uncontrolled_uprime_18_26T}
\end{figure}

Figure \ref{fig: uncontrolled_uprime_18_26T} shows contour plots of the instantaneous streamwise velocity fluctuation in x-y plane. Streamwise elongated high and low speed streaks, resulting from the penetration of the low frequency inlet perturbation inside the boundary layer, can be clearly seen. The streamwise wavelength of these streaks is approximately $740$ units and analysis show that they convect downstream at $0.75U_{\infty}$. These values are in agreement with the temporal frequency of the penetrating inlet mode A ($\omega=0.0064$). The streamwise wavelength of the undulations seen at top of the boundary layer is $50$ units and their convection speed is $U_{\infty}$. The resulting temporal frequency agrees with the high frequency inlet mode B ($\omega=0.128$). The time difference between the two plots is $\Delta t=480$, which is equivalent to half period of mode A. During this time, the positive streak at $x=400$ in the top plot, propagates downstream a distance of half wavelength. Negative streaks are lifted upwards to the top of the boundary layer, where they interact with the high frequency disturbance and break down. This leads to secondary instability and the inception of turbulent spots. Such an instability can be seen in the top figure, between $x \approx 800-1000$ and in the middle of the boundary layer, $y \approx 5-15$. Once spots form, they grow in all directions and impinge on the wall (top-bottom effect). For example, the aforementioned instability has propagated in the x direction and has impinged on the wall, as shown in the bottom figure. Similar observations were made in the simulations of \cite{jacobs2001simulations}, \cite{zaki2005mode}, \cite{brandt2004transition}. This type of secondary instability is known as outer mode and has been studied by \cite{andersson2001breakdown, vaughan2011stability}. The high-speed streaks on the other hand are not affected by the external perturbations and they stay close to the wall. One such high speed streak is shown to enter the region of the control slot in the top figure; it is located between $x \approx 900-1100$ and $y \approx 0-3$. \textcolor{black}{Note that the interaction with high frequency mode occurs upstream of the control slot and therefore in the controlled flow, the transition is also via the outer mode. As can be seen in figure \ref{fig: uncontrolled_uprime_18_26T}, the controller is located in the region of streak breakdown where the flow is fully nonlinear.}

\begin{figure}[!htbp]\centering
\includegraphics[width=0.7\textwidth]{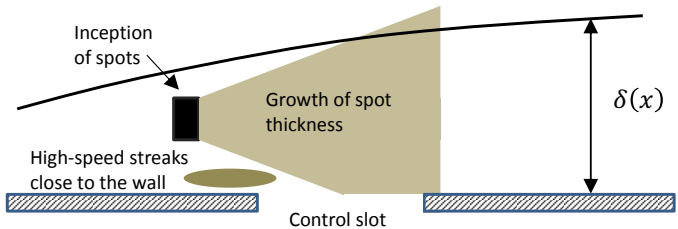}
\caption{Growth of spots and impingement on the control region.}
\label{fig:spot_growth}
\end{figure}

A sketch that summarises the breakdown process due to the outer instability is shown in figure \ref{fig:spot_growth}. This sketch will be refered to later when the control action is characterised.

\subsection{Nonlinear optimal control algorithm}
The objective of the nonlinear optimal control algorithm is to find the actuation velocity that locally minimises a cost function, within a finite optimisation time $T$. The actuation method is blowing and suction at the wall, which varies in time and space. The location of the control slot is in the late transition zone, from $x=1050$ to $1300$ (Figure \ref{fig: uncontrolled_uprime_18_26T}). The cost function is defined as,
\begin{equation}
\label{eq: Cost function}
\mathcal{J}=\int_0^T E(t)dt +l^2\int_0^T E_w(t)dt,
\end{equation}
where $T$ is the length of optimisation time and $l^2$ is a weighting parameter that penalise the magnitude of actuation. For example, a small value of $l^2$ indicates low penalization and results in higher control velocities. The first term on the right hand side of the cost function (\ref{eq: Cost function}) is the energy of the flow, which is defined as a quadratic measure of the deviation of the instantaneous velocity $\textbf{u}(t)$ from the Blasius velocity profile $\textbf{U}_{B}$,
\begin{equation}
E(t)=\int_V\left[(\textbf{u}(t)-\textbf{U}_{B})\cdot \Omega \left(\textbf{x}\right) \cdot (\textbf{u}(t)-\textbf{U}_{B})\right]dV,
\label{eq:flow_energy}
\end{equation}
\noindent where $\Omega \left(\textbf{x}\right)$ is an indicator function that specifies a subregion ($1050<x<1350$, $0<y<5$, $0<z<90$) inside which the flow energy is to be reduced ($\Omega(\textbf{x})=1$ inside and $\Omega(\textbf{x})=0$ outside). For simplicity, in the rest part of this paper, $E$ is called flow energy. Therefore the objective of the control algorithm is to drive the velocity field towards the Blasius solution. The analysis of the control performance is also easier since the target profile is known.  

The second term on the right hand side of the cost function (\ref{eq: Cost function}) measures the cost of the control and is defined as,
\begin{equation}
E_w (t) =\int_{{S_w}} \left[v_w(t)\right]^2dS_w,
\end{equation}
where $v_w$ is the actuation velocity and ${S_w}$ is the area of the wall, including the control slot (of course outside the slot $v_w=0$). 

The cost function is to be minimised while all the constraints describing the flow problem (e.g. governing equations, initial conditions) are satisfied. Using the Lagrange multiplier technique, the constrained flow problem is replaced by an unconstrained one defined by the Lagrange cost function $\mathcal{L}$,
\begin{equation}
\mathcal{L}=\mathcal{J}-\textbf{a}\cdot \textbf{F},
\end{equation}
where $\textbf{a}$ is the Lagrange multiplier, $\textbf{F}$ includes all the constraints, and $\textbf{a}\cdot \textbf{F}$ denotes the dot product operation. The problem now reduces to finding the flow variables ($\textbf{u}$ and $p$) and the Lagrange multipliers ($\textbf{u}^{\dagger}$ and $p^{\dagger}$) that minimise $\mathcal{L}$.  A differentiate-then-discretize approach was used. The first order variation of the Lagrangian with respect to each independent argument must be zero at a stationary point. Setting the first variation of $\mathcal{L}$ with respect to $\textbf{u}(u,v,w)$ and $p$ to zero and after integration by parts, the following linear system of adjoint equations is derived,
\begin{equation}
  \nabla \cdot \textbf{u}^{\dagger} =  0,
 \end{equation}
 \begin{equation}\label{eq: adjoint equation}
  \frac{\partial \textbf{u}^{\dagger}}{\partial t}=-\left(\textbf{u}\cdot \nabla \right)\textbf{u}^{\dagger}+\textbf{u}^{\dagger}\cdot \left(\nabla\textbf{u}^{T}\right)-\nabla p^{\dagger}-\frac{1}{Re_{L_0}}\nabla^2\textbf{u}^{\dagger}+2\left(\textbf{u}-\textbf{U}_B \right)\Omega,
\label{eq: adjoint continuity equation}
\end{equation}
where $\textbf{u}^{\dagger}$ and $p^{\dagger}$ are the adjoint velocity and pressure, respectively. Note that the above system of linear equations is solved by integrating backwards in time. The terminal conditions for the adjoint variables are,
\begin{equation}
\textbf{u}^{\dagger}(T)=\textbf{v}^{\dagger}(T)=\textbf{w}^{\dagger}(T)=0.
\end{equation}
The boundary conditions are provided in \cite{xiao2017nonlinear}. The Navier-Stokes equation and the adjoint equation form a nonlinear coupled system, which is solved iteratively. The first variation of $\mathcal{L}$ with respect to the control input $v_w$ is used to update the control velocity at each iteration,
\begin{equation}\label{eq: actuation velocity update}
v_w^{n+1}(x,z,t)=v_w^n(x,z,t)-\alpha^n\left(\frac{\partial \mathcal{L}}{\partial v_w(x,z,t)}\right),
\end{equation}
where $n$ is the iteration number and $\alpha$ is an adjustable step length.

The optimization procedure for a finite time $T$ is summarised as follows,
\begin{enumerate}
\item Assume an initial distribution for the actuation velocity $v_w^0(x,z,t)$ (usually 0) and an initial value for $\alpha$; 
\item Solve the Navier-Stokes equations forward from $t=0$ to $t=T$; \label{item:2}
\item At $t=T$, evaluate the values of the objective function (\ref{eq: Cost function}) between two successive iterations.
\begin{itemize}
\item If $\mathcal{J}$ decreases, and the change is smaller than a predefined threshold, stop iteration loop, otherwise continue to step (\ref{item:4});
\item If $\mathcal{J}$ increases, halve the value of $\alpha$ and continue to step (\ref{item:4});
\end{itemize}
\item Integrate the adjoint equation backward from $t=T$ to $t=0$; \label{item:4}
\item At $t=0$, update the control velocity using (\ref{eq: actuation velocity update}) and return to step (\ref{item:2}).
\end{enumerate}

\begin{figure}\centerline{
\includegraphics[width=0.99\textwidth]{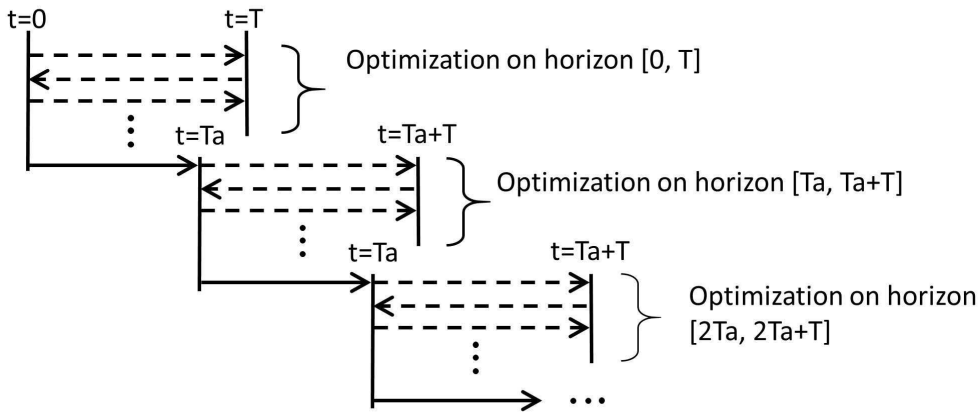}}
\caption{\label{fig: receding_horizon} Schematic representation of the receding horizon framework.}
\end{figure}

\citet{xiao2017nonlinear} performed simulations with three optimization lengths $T$ and found that the larger $T$ results in a larger flow energy reduction. As mentioned in the introduction, the maximum value of $T$ is limited by the instability of the adjoint equations when integrating backwards in time. To perform control simulations for longer time, the optimization approach is applied in a receding horizon framework. A schematic representation of the procedure is shown in figure \ref{fig: receding_horizon}. Once convergence is achieved within $T$, the flow is advanced in time by some portion $T_a$ of $T$ and a new optimisation problem is solved again. The actuation is optimal over $T$ and sub-optimal during the whole multi-T period examined. As \citet{bewley2001dns} correctly pointed out, the actuation computed near the end of each optimisation interval is determined without considering further development of the flow, as opposed to the actuation obtained near the beginning of each interval. Therefore the actuation near the end of $T$ may not be as effective as the one at the beginning of $T$. 

In the present simulations we \textcolor{black}{used the longest $T=60$ for which the adjoint variables provide reliable sensitivities.} Due to the high computational cost required for the receding horizon method, we take $T_a=T$ in order to maximise the total time span. The weighting parameter $l^2$ in the objective function (\ref{eq: Cost function}) is set to $l^2=150$ so that the mean actuation velocity is less than $1\% U_{\infty}$. The maximum value of the adjustable step length is set to be $\alpha=0.001$ to ensure smooth convergence.

\section{Results}\label{sec:results}
\subsection{Flow energy}\label{sec: Flow energy}
A total of 48 optimisation intervals, corresponding to a time duration equal to $48T=2880$, are considered. This is approximately equal to three complete periods of the low-frequency, penetrating inlet mode A (table \ref{tab: parameters for two modes}). This time span is long enough for the controlled flow to reach the exit of the computational domain. 

\begin{figure}\centerline{
\includegraphics[width=0.9\textwidth]{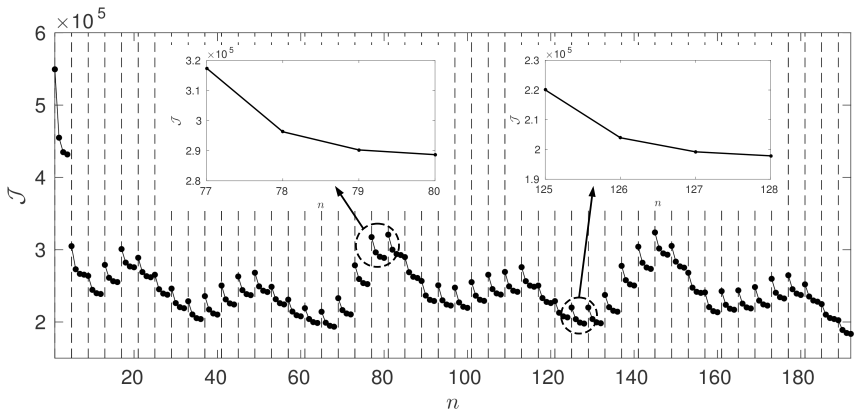}}
\caption{\label{fig: cost_function} Cost function $\mathcal{J}$ versus the total number of iterations n. The vertical dashed lines delineate the optimization intervals. The two inset figures show the convergence of the cost function in the $20^{th}$ and $32^{nd}$ interval.}
\end{figure}

The convergence of the cost function $\mathcal{J}$ (defined in \ref{eq: Cost function}) within each optimization interval is shown in figure \ref{fig: cost_function}. Recall that $\mathcal{J}$ is defined as the sum of the integral of flow energy over time $T$ plus the cost of actuation. Due to the high computational cost required, the maximum number of iterations in each interval was set to 4. As figure \ref{fig: cost_function} clearly demonstrates, there is a large drop of $\mathcal{J}$ in the first optimisation interval $T$ because the actuation starts from an uncontrolled transitional flow state. After the second interval, $\mathcal{J}$ exhibits a repeatable oscillatory pattern, with frequency equal to that of the low frequency inlet mode A (indeed 3 cycles can be detected). The shape of the oscillatory pattern is related to the transition activity taking place inside the control region and will be explored in more detail below. Two examples of convergence of the cost function can be seen in the inset figures; both indicate smooth convergence. The reduction of $\mathcal{J}$ between the 3rd and 4th iteration is small, so adding further iterations would increase computational cost, without commensurate reduction of the cost function. The low frequency penetrating mode also modulates the flow, again making more iterations within a single $T$ unnecessary.

\begin{figure}\centerline{
\includegraphics[trim={0cm 0cm 0cm 0cm},clip=true,width=0.9\textwidth]{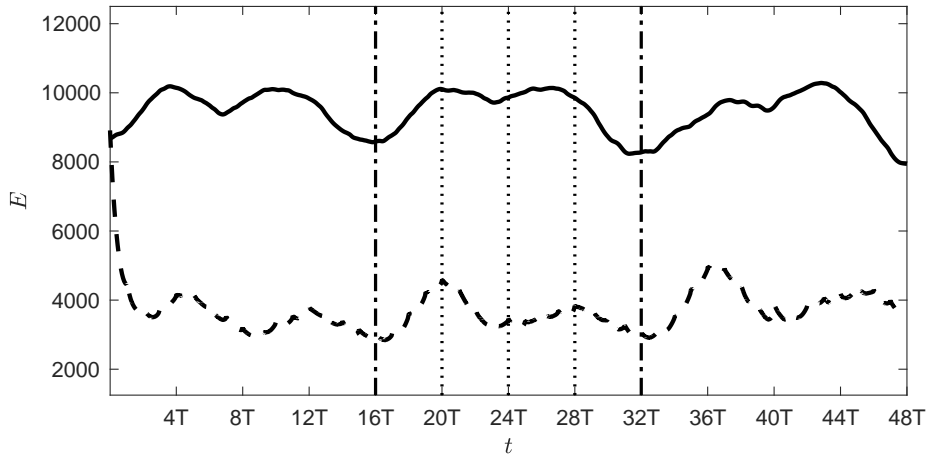}}
\caption{Evolution of the energy of the flow defined in the defined subregion. Solid line, uncontrolled flow; dashed line, controlled flow.}
\label{fig: Energy_flow} 
\end{figure}

The time evolution of the flow energy $E$ for both uncontrolled and controlled flow is shown in figure \ref{fig: Energy_flow}. Note that in the horizontal axis, time is expressed in terms of the optimization length $T$ (with $T=60$). For the uncontrolled flow, three complete periods of the low frequency mode ($0-16T$, $16-32T$ and $32-48T$) can be detected from the shape of $E$; these are delineated by 2 vertical dash-dot lines. It is clear that the breakdown process that determines $E$ is modulated by the low frequency external mode A. \textcolor{black}{The first few intervals (the very first interval is exactly the same as in \citet{xiao2017nonlinear}) can be regarded as the initial transient period until the control effect is established after $16T$. In the current work, we mainly focus on the control effect and the associated control mechanism in the time domain $16T-48T$, during which the controlled flow is very different  compared to the one during the first few intervals.}

In figure \ref{fig: Energy_flow}, the controlled flow energy is reduced significantly within the first few optimisation intervals, and then stays in the range between $3000-4500$. This is approximately $43\%$ of the uncontrolled flow energy. Between $0-16T$, there is a small phase difference between controlled and uncontrolled flow. This is because when the controller is activated at $t=0$, the uncontrolled flow is in a state of increasing $E$. From $t=16T$ onwards, the two energies evolve in phase. Similar flow processes take place in the controlled flow, but with reduced flow energy.

\begin{figure}\centering
\sidesubfloat[]{
\includegraphics[trim={0cm -0.2cm 0cm -0.2cm}, clip=true,width=0.48\textwidth]{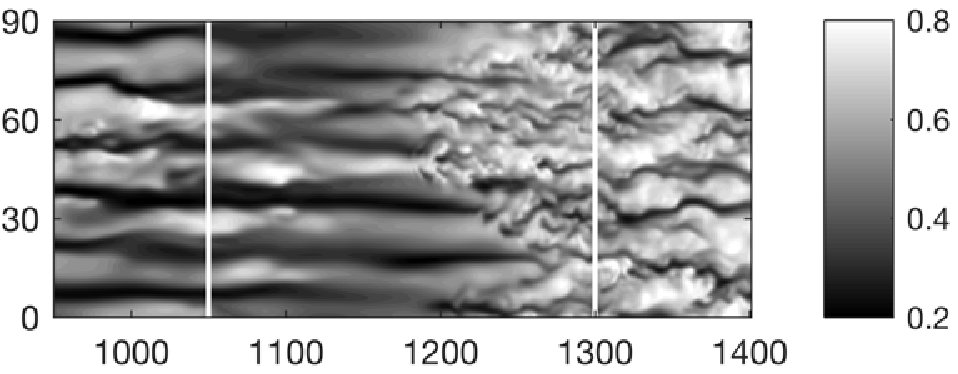}}
\sidesubfloat[]{
\includegraphics[trim={0cm -0.2cm 0cm -0.2cm}, clip=true,width=0.48\textwidth]{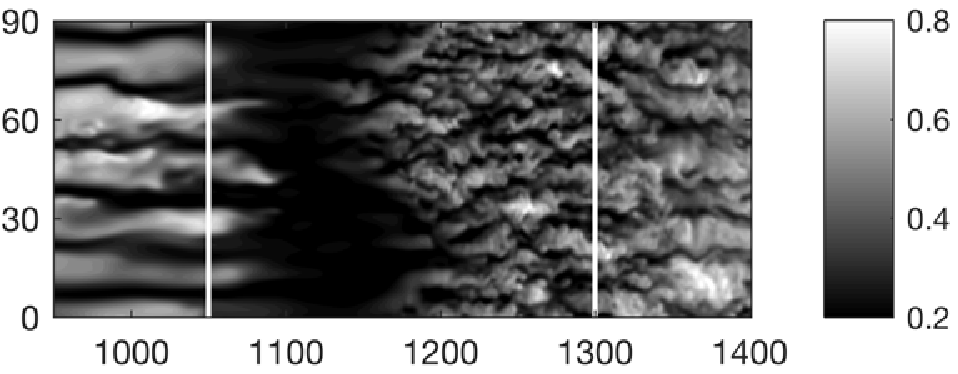}}\quad
\sidesubfloat[]{
\includegraphics[trim={0cm -0.2cm 0cm -0.2cm}, clip=true,width=0.48\textwidth]{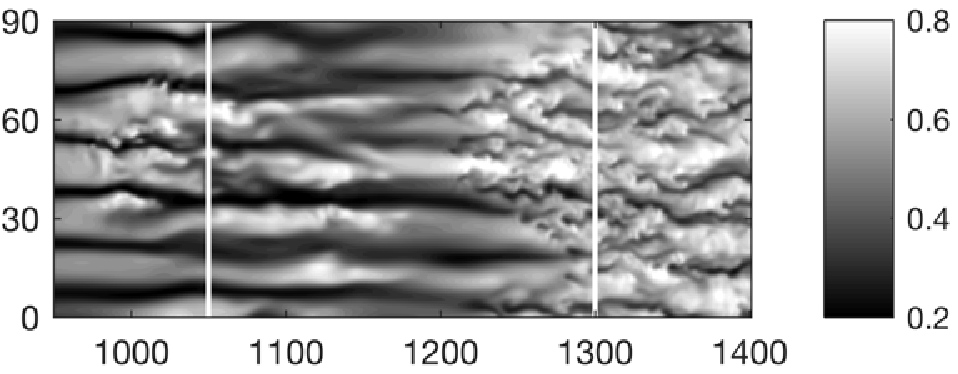}}
\sidesubfloat[]{
\includegraphics[trim={0cm -0.2cm 0cm -0.2cm}, clip=true,width=0.48\textwidth]{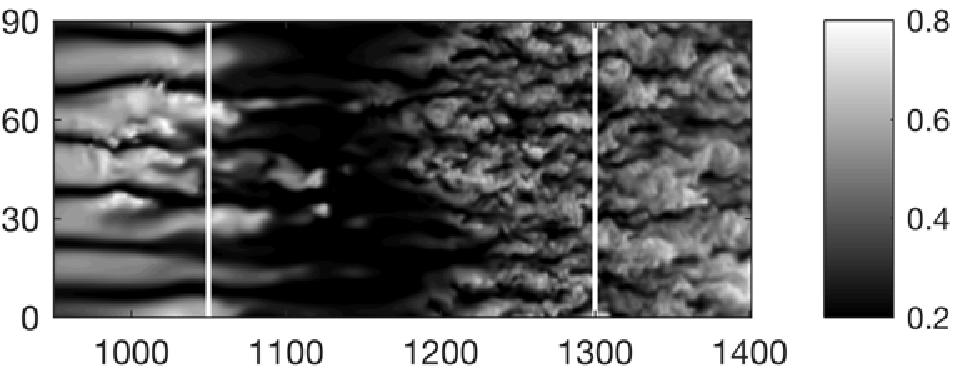}}\quad
\sidesubfloat[]{
\includegraphics[trim={0cm -0.2cm 0cm -0.2cm}, clip=true,width=0.48\textwidth]{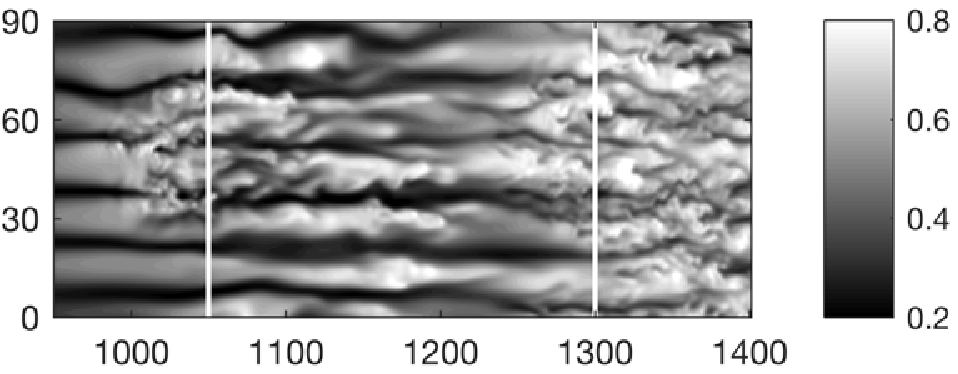}}
\sidesubfloat[]{
\includegraphics[trim={0cm -0.2cm 0cm -0.2cm}, clip=true,width=0.48\textwidth]{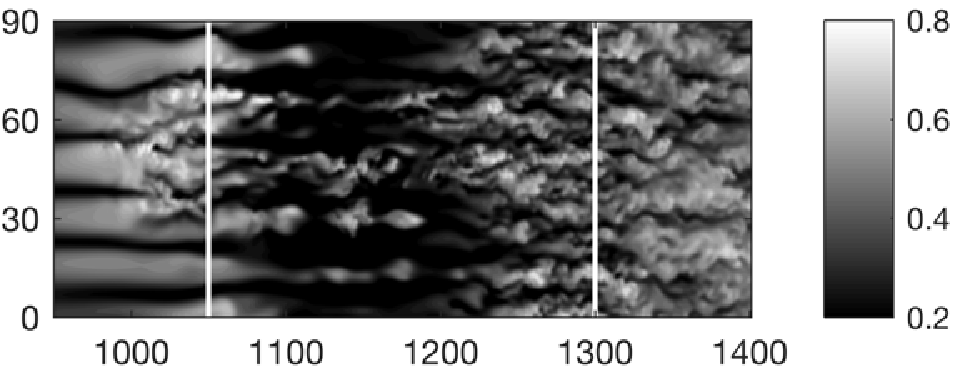}}\quad
\sidesubfloat[]{
\includegraphics[trim={0cm -0.2cm 0cm -0.2cm}, clip=true,width=0.48\textwidth]{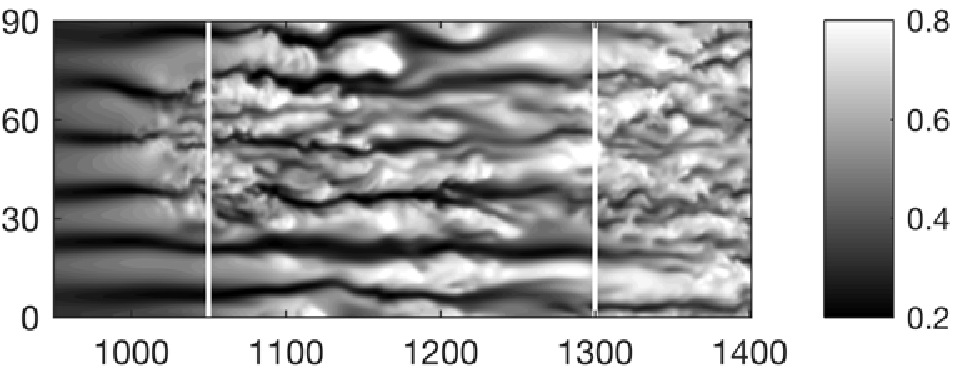}}
\sidesubfloat[]{
\includegraphics[trim={0cm -0.2cm 0cm -0.2cm}, clip=true,width=0.48\textwidth]{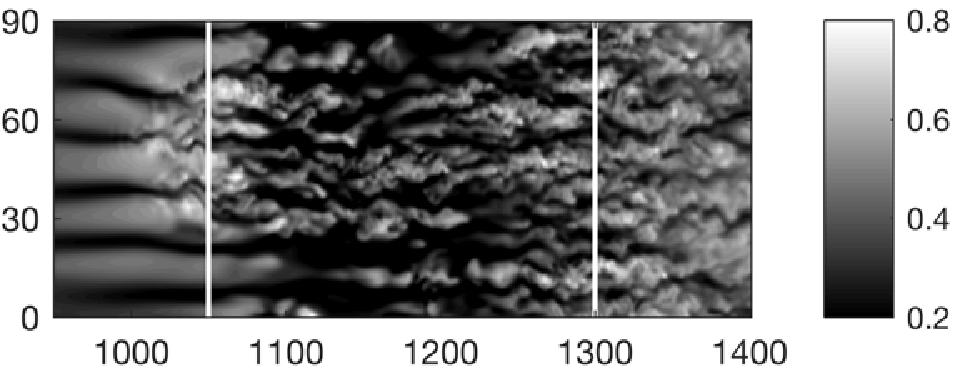}}
\caption{Uncontrolled (left) and controlled (right) instantaneous streamwise velocity in x-z plane at $y=2$ at four successive time instants: (a), (b) 16T; (c), (d) 17T; (e), (f) 18T; (g), (h) 19T.}
\label{fig: u_16_19T}
\end{figure}

The evolution of $E$ indicates that two local maxima appear within each period.The evolution of the uncontrolled and controlled flow is examined by visualising the instantaneous $u$ in figure \ref{fig: u_16_19T} at 4 successive time instants. The total duration corresponds to slightly less that one quarter of the period of the low frequency mode (the vertical dotted lines in figure \ref{fig: Energy_flow} delineate each quarter). Figure \ref{fig: u_16_19T} demonstrates in detail the evolution of one patch of streaks as it convects into the control region. {In the uncontrolled flow at $t=16T$, the head of one patch of high-speed streaks is seen to enter the control slot, while the downstream side is occupied by turbulent flow. Secondary instabilities distort the streaks entering the slot. The front part convects more rapidly than the rear and soon overtakes the downstream turbulent region. The ragged edge of the turbulent region is maintained by turbulent spots that continuously overtake the main turbulent zone. Without this sustenance the turbulent flow would convect out of the numerical domain \citep{jacobs2001simulations}. In figure \ref{fig: Energy_flow} it can be seen that $E$ is at a local minimum at $t=16T$. This is due to the region of inactivity between the entering high-speed streaks and the turbulent spot. At $t=19T$, the actuation region is almost full of turbulent flow, and the flow energy is approaching a local peak value, as expected.} 

In the controlled flow, at all time instants, the high-speed streaks are still visible and distorted, \textcolor{black}{indicating that the penetrating mode is present}, but they have been clearly quenched by the control action, while the turbulent zone has also been significantly  affected. Between them, there is extended region with very low velocities and small spanwise variation in $u$. In the downstream end, there are some small localised patches of large $u$ but they appear random. This figure indicates that the controller has been very effective in suppressing the energy of the flow. It is very interesting to notice that the control effect has propagated downstream of the control slot. This aspect of control action will be further investigated in section \ref{sec: control effect downstream of the slot}. 

\begin{figure}\centering
\includegraphics[trim={0cm 0cm 0cm 0cm},clip=true,width=0.9\textwidth]{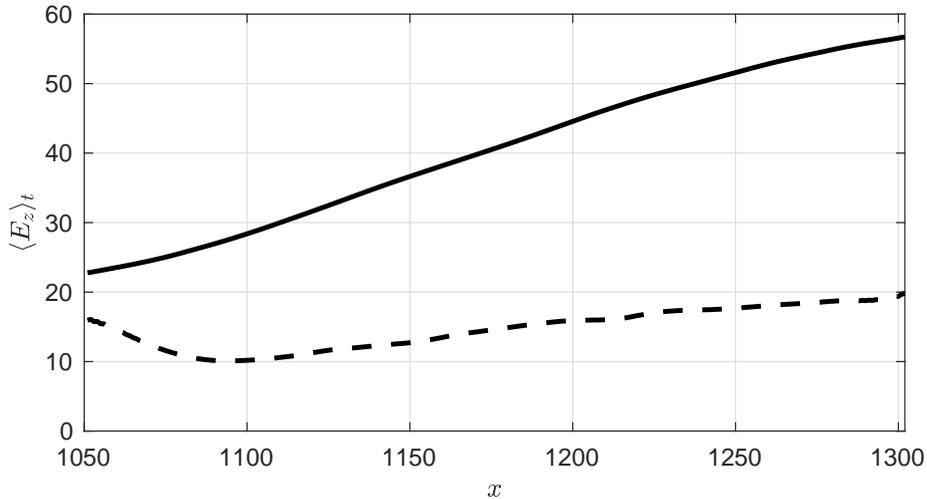}
\caption{Streamwise variation of $\langle E_z \rangle_t$. Solid line, uncontrolled flow; dashed line, controlled flow.}
\label{fig: Energy_flow_spatial}
\end{figure}

To further understand the control effect, we define $E_z(x,t)$ as the energy integrated in the spanwise and wall-normal directions,
\begin{equation}\label{eq: spanwise disturbance energy}
E_z(x,t)=\int_0^{y_{lim}}\int_0^{L_z} \left[(\textbf{u}(t)-\textbf{U}_{B})\cdot \Omega \left(\textbf{x}\right) \cdot (\textbf{u}(t)-\textbf{U}_{B})\right]dydz
\end{equation}
The time averaged value $\langle E_z \rangle_t$ is plotted in figure \ref{fig: Energy_flow_spatial} against $x$. From the definition of $E_z(x,t)$, the area under the curve is equal to the total energy of the flow above the slot. In the uncontrolled case, the energy increases almost linearly as a result of the transition process taking place over the control slot, as described above. In the controlled flow, $\langle E_z \rangle _{t}$ is already reduced at $x=1050$ (starting position of the control slot) compared to the uncontrolled flow. This indicates that, in a time-average sense, the actuation effect is already felt upstream of the slot. The controlled $\langle E_z \rangle _{t}$ then decreases and reaches a minimum value just before $x=1100$. After this location, it increases linearly, but with a much smaller rate compared to the uncontrolled flow. This behaviour is in agreement with the $u$ velocity contours of figure \ref{fig: u_16_19T}. 

\subsection{Optimal control velocity}\label{sec: optimal control velocity}

\begin{figure}[!htbp]\centering
\includegraphics[trim={0cm 0cm 0cm 0cm},clip=true,width=0.9\textwidth]{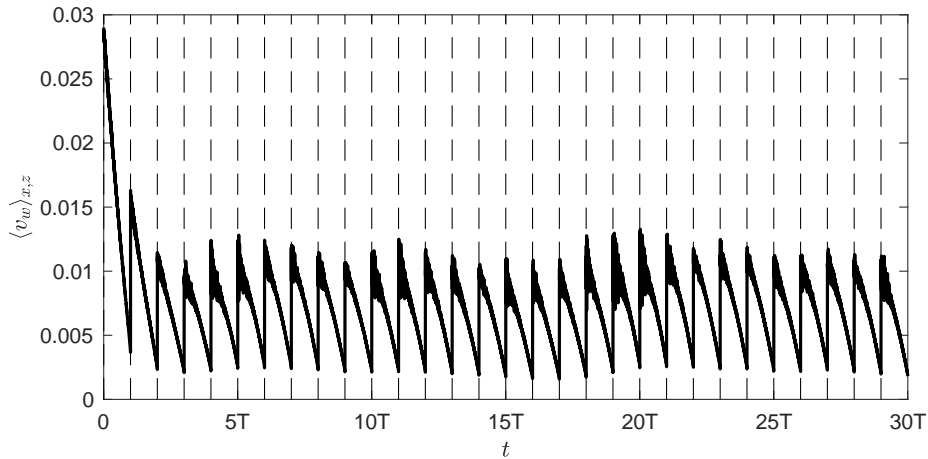}
\caption{\label{fig: vw_mean} Spatially averaged actuation velocity against time. Vertical dashed lines demarcate optimisation intervals.}
\end{figure}

The spatially (streamwise and spanwise) averaged optimal blowing and suction velocity $\langle v_w \rangle_{x,z}$ is shown in figure \ref{fig: vw_mean} as a function of time. Only the first 30 optimisation intervals are shown; the rest have very similar behaviour. The actuation velocity $\langle v_w \rangle_{x,z}$ attains its maximum value at the beginning of each interval and gradually reduces towards the end, which is in agreement with the actuation from control in a single optimisation horizon \citep{xiao2017nonlinear}. After a short transient period at the beginning of actuation, the value of $\langle v_w \rangle_{x,z}$ within each interval stabilises to an average of $0.685\% U_{\infty}$. It was demonstrated in figure \ref{fig: Energy_flow} that the flow energy is reduced dramatically in the first few intervals and then fluctuates around a lower value. The control velocity has a similar behaviour. This suggests that a larger actuation velocity is required to bring down the uncontrolled flow at the beginning of actuation,  and subsequently a smaller $v_w$ is enough to maintain the flow energy at that lower level. 

\begin{figure}[!htbp]\centering
\includegraphics[trim={0cm 0cm 0cm 0cm},clip=true,width=0.9\textwidth]{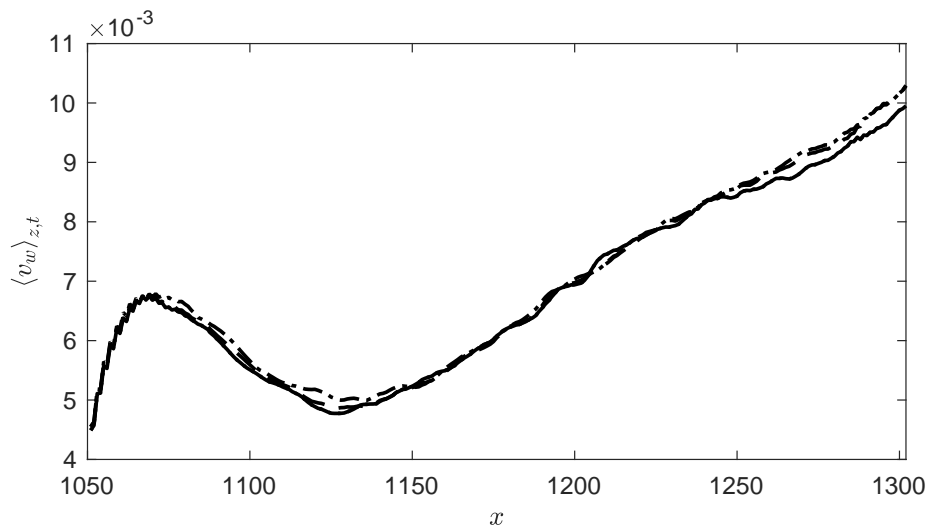}
\caption{Time- and spanwise- averaged actuation velocity. The 4 distributions shown are averaged over one period of low frequency inlet mode (i.e. 16T) starting from different time instants: solid line, 6T; dashed line, 18T; dash-dot line, 22T; dotted line, 26T.}
\label{fig: spanwise_averaged_vw_period}
\end{figure}

In figure \ref{fig: spanwise_averaged_vw_period} the variation of the time- and spanwise- averaged actuation velocity $\langle v_w \rangle_{z,t}$ with streamwise distance inside the control slot is plotted. Time averaging here is performed over one period of the low frequency inlet mode A (i.e. over $16T$), starting from 4 different time instants (mentioned in the figure caption). All four curves collapse reasonably well. This indicates that the spanwise averaged actuation velocity is repeatable in a time scale equal to the period of the low frequency inlet mode. We noticed in figure \ref{fig: Energy_flow} that the flow energy is periodically modulated by the slow mode, so it is not surprising that $\langle v_w \rangle_{z,t}$ also demonstrates periodic behaviour.  Note also that when no constraint on mass is imposed, the control velocity results in a net positive mass flow rate.

It is interesting to examine the streamwise variation of $\langle v_w \rangle_{z,t}$. When the optimisation was performed on a single interval \citep{xiao2017nonlinear}, $\langle v_w \rangle_{z}$ increased \textcolor{black}{monotonically} in the streamwise direction and a physical explanation was provided based on the flow pattern and the definition of the objective function. However when the actuation velocity is averaged over much longer time, there is a local peak at around $1070$, a local minimum at 1130, and $\langle v_w \rangle_{z,t}$ then increases linearly with $x$. 

This behaviour is consistent with the instantaneous flow patterns of the controlled flow (right column of figure \ref{fig: u_16_19T}). The peak at $1070$ corresponds to the control action to quench the incoming distorted high-speed streaks. The minimum value corresponds to the low velocity region (black area), and the linear growth to the suppression of the turbulent patches with large $u$ that appear in the right half of the slot, but they originate upstream (refer to figure \ref{fig:spot_growth}).

It is interesting to notice that the profiles of both the controlled energy $\langle E_z \rangle_t$ (figure \ref{fig: Energy_flow_spatial}) as well as the actuation $\langle v_w \rangle_{z,t}$ (figure \ref{fig: spanwise_averaged_vw_period}) reduce due to the control action in the streamwise direction, but then rise again. This behaviour is the result of the competition between the control action and the transition process. If transition to turbulence had not taken place above the control slot, the flow energy as well as $\langle v_w \rangle_{z,t}$ would continue to reduce, in a manner similar to that found by \citet{monokrousos2008dns}. In their work, the control slot is placed in the upstream laminar flow region and both the flow and control velocity reduce \textcolor{black}{monotonically} in the streamwise direction within the slot. This is the expected behaviour if there is no transition.

In the present case however, the interaction between the elevated negative streaks and the external high frequency mode that leads to secondary instability, starts inside the boundary layer and upstream of the control slot as seen in figure \ref{fig: uncontrolled_uprime_18_26T}. \cite{nolan2013conditional} used laminar-turbulent discrimination techniques and managed to detect the inception and growth of spots. They showed that the highest turbulent spot count appears inside the boundary layer, in the region $y/\delta = 0.4-0.8$. The current objective function minimises the flow energy closer to the wall, below $y/\delta_c=0.25$, where $\delta_c$ is the average boundary layer thickness over the control slot (see also top plot in figure \ref{fig: shape_factor}). The spots are then brought towards the wall, and the control velocity in response to this, increases along $x$ as seen in figure \ref{fig: spanwise_averaged_vw_period}. This top-to-bottom process in relation to the control slot was sketched in figure \ref{fig:spot_growth}. 

In the following two sections, the effect of control action on the time-average flow above and downstream of the slot until the end of the domain is examined. For a more accurate and meaningful comparison, all the mean flow properties for both controlled and uncontrolled flow are obtained by averaging over two periods of low frequency mode A, starting from $t=16T$ until $t=48T$. The reason is that at $t=16T$ the effect of actuation has reached the end of the domain (as will be shown later), so the same time window can be used for the flow both above and downstream of the slot.

\subsection{Effect of actuation on mean and turbulent quantities above the control slot}\label{sec: control effect inside the controller region}

\subsubsection{Mean velocity profiles}\label{subsec: mean velocity profiles}
\begin{figure}\centering
\sidesubfloat[]{
\includegraphics[trim={-0.cm -0.cm -0.30cm -0.cm}, clip=true,width=0.28\textwidth]{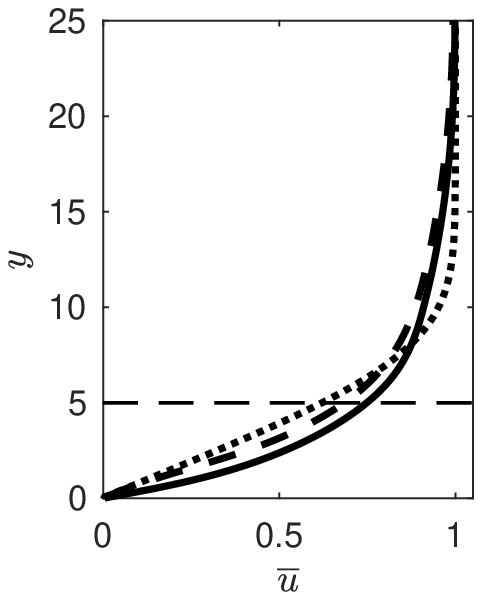}}
\sidesubfloat[]{
\includegraphics[trim={-0.cm -0.cm -0.30cm -0.cm}, clip=true,width=0.28\textwidth]{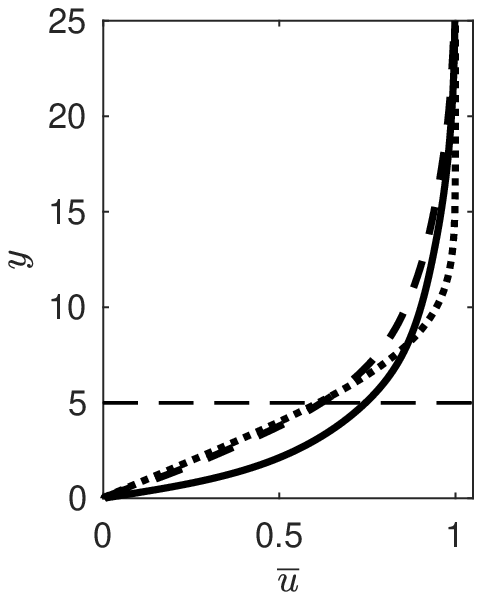}}\sidesubfloat[]{
\includegraphics[trim={-0.cm -0.cm -0.30cm -0.cm}, clip=true,width=0.28\textwidth]{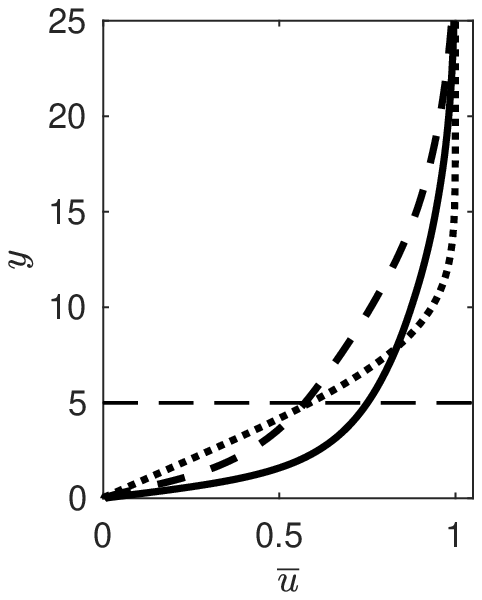}}
\caption{\label{fig: mean_v_three_locations} Time- and spanwise averaged velocity profiles at three locations: (a) $x=1075$; (b) $x=1130$; (c) $x=1250$. Solid line, uncontrolled flow; dashed line, controlled flow; dotted line, Blasius profile. The horizontal dashed line marks the upper boundary in the wall-normal direction in which the cost function is defined.}
\end{figure}

The mean (time- and spanwise- averaged) velocity profiles $\overline{u}$ at three streamwise locations ,$x=1075$, $1130$ and $1250$, are shown in figure \ref{fig: mean_v_three_locations}. These locations are selected based on the previous analysis on the streamwise variation of the optimal control velocity. They correspond to the peak of actuation close to the inlet, the trough in the quiescent region between steaks and turbulent flow and to a location inside the fully developed region, respectively. 

It can be seen clearly that the control has imparted significant changes in the mean profiles in the three locations. At $x=1075$, the controlled velocity profile is between the uncontrolled and Blasius profile, while at $x=1130$, the controlled flow matches well with the Blasius solution.  Further downstream at $x=1250$, the uncontrolled profile is steeper as a result of the enhanced mixing arising from the transition process. The controlled flow velocity has milder slope, but still larger than the Blasius one, indicating that the controlled flow is still disorganised, but with reduced mixing activity. This is in agreement with the findings of the previous section. As already mentioned, the upper boundary of the box in which the objective function is defined is at $y=5$; this position is indicated by a horizontal dashed line in figure \ref{fig: mean_v_three_locations}. The optimal controller performs its "duty" as expected; it brings the velocity profile close to the Blasius velocity only below $y=5$. 

The optimal actuation velocity depends on the choice of the objective function. If the location of the upper boundary were set closer to the wall, one would expect that the controlled velocity profile would approach even better the Blasius profile in the near wall region, leading to lower drag (closer to laminar drag). It is therefore clear that minimizing energy is not expected to lead to drag minimization if the upper boundary is set far from the wall.

\begin{figure}[!htbp]\centering
\includegraphics[trim={0cm 0cm 0cm 0cm},clip=true,width=0.8\textwidth]{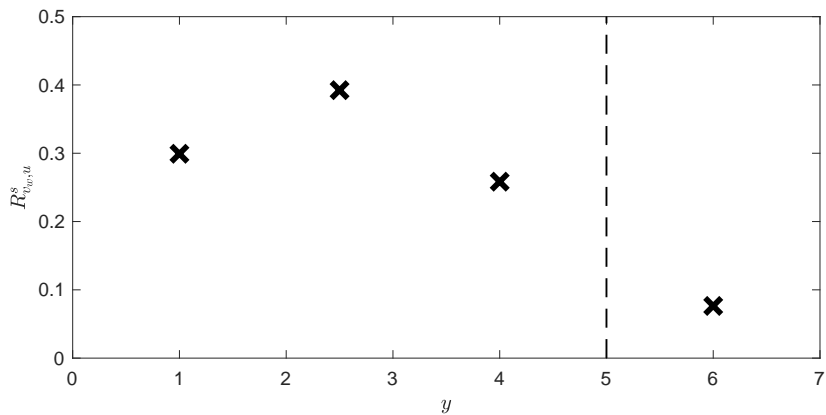}
\caption{\label{fig: correlation_space_y} Time-averaged (over one period) correlation coefficient between control velocity over the whole control region and instantaneous streamwise velocity at various wall-normal locations over the whole control region.}
\end{figure}

The spatial correlation between $v_w$ and $u$, $R_{v_w,u}^{s}$, evaluated over the whole control slot is now examined. $R_{v_w,u}^{s}$ is defined as,
\begin{equation}
R_{v_w,u}^{s}=\frac{\langle \left(v_w-\langle v_w\rangle_{x,z} \right)\left(u-\langle u \rangle_{x,z}\right)\rangle_{x,z}}{\left[\langle \left(v_w-\langle v_w\rangle_{x,z} \right)^2\rangle_{x,z} \langle \left(u-\langle u \rangle_{x,z} \right)^2\rangle_{x,z}\right]^{1/2}}
\end{equation}
where $\langle \rangle_{x,z}$ denotes average in streamwise and spanwise directions. Figure \ref{fig: correlation_space_y} shows $R_{v_w,u}^{s}$ averaged over one period of time with $u$ extracted from four wall-normal locations. It can be seen that $R_{v_w,u}^{s}$ peaks at $y=2.5$, which means $v_w$ is most responsive to the flow in this region. The explanation for this is that the difference between the uncontrolled $\overline{u}$ and Blasius velocity is larger between $y=2-3$ as seen in figure \ref{fig: mean_v_three_locations}. Note also that  $v_w$ is very weakly correlated with $u$ at $y=6$. This is not surprising since this location is outside the region where the objective function is defined and the controller does not respond to the flow in this region, therefore $R_{v_w,u}^{s}$ is very small. Above $y=5$, the controlled flow profile deviates strongly from Blasius profile compared to the uncontrolled flow. This again ties with the low correlation between $v_w$ and $u$ at $y=6$.

\begin{figure}\centering
\sidesubfloat[]{
\includegraphics[trim={-1.cm -0.cm -0.cm -0.cm}, clip=true,width=0.82\textwidth]{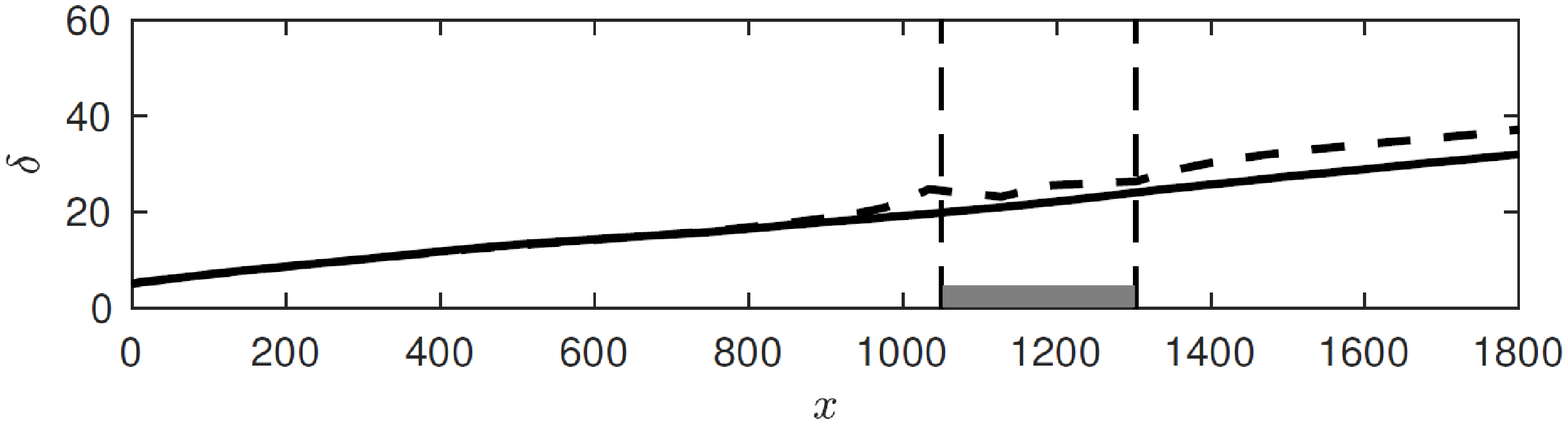}}\quad
\sidesubfloat[]{
\includegraphics[trim={0.cm -0.cm -0.0cm -0.cm}, clip=true,width=0.8\textwidth]{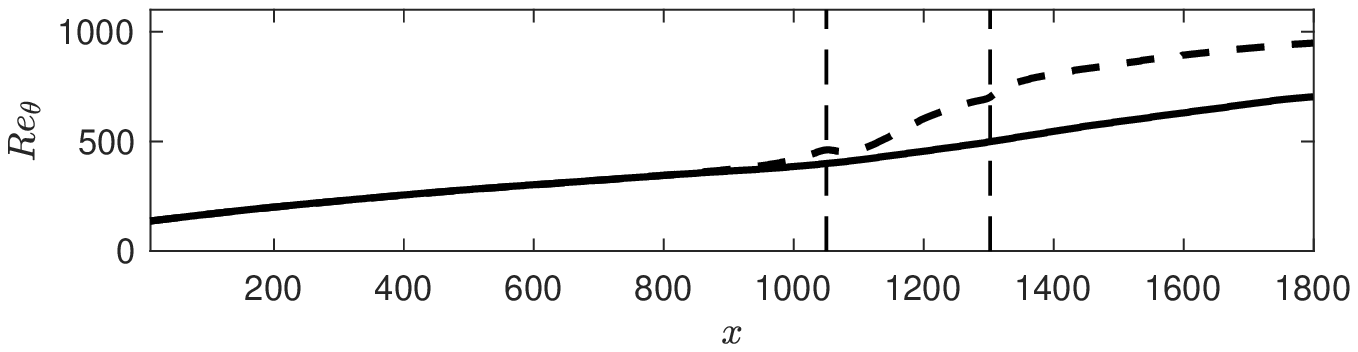}}\quad
\sidesubfloat[]{
\includegraphics[trim={-0.35cm -0.cm -0.cm -0.cm}, clip=true,width=0.8\textwidth]{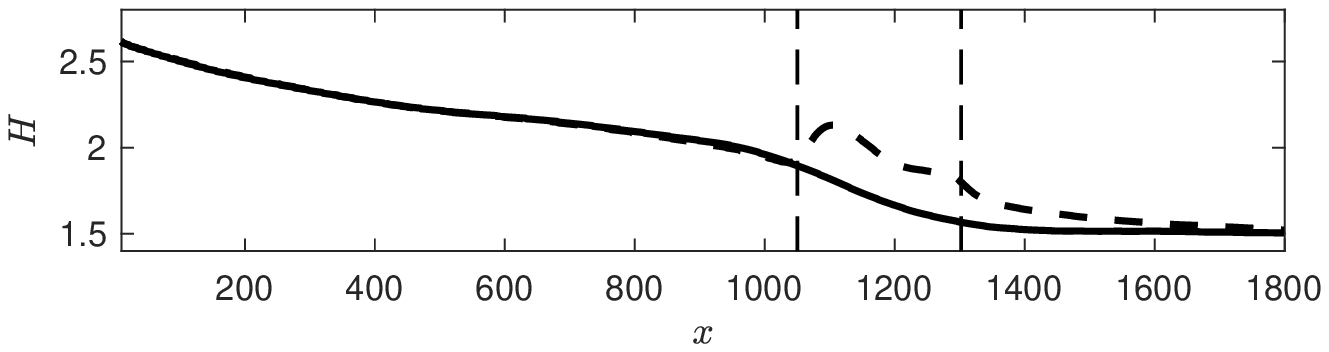}}
\caption{\label{fig: shape_factor} Effect of optimal control on spatial development of the boundary layer: top, boundary layer thickness $\delta$; middle, Reynolds number based on momentum thickness $Re_{\theta}$; bottom, shape factor $H$. Solid line, uncontrolled flow; dashed line, controlled flow. In (a) the shaded area denotes the region where the cost function is defined.}
\end{figure}

The effect of the control action on the spatial development of the boundary layer is further investigated by examining the boundary layer thickness $\delta$, the Reynolds number based on momentum thickness $Re_{\theta}$ and the shape factor $H$, which are plotted in figure \ref{fig: shape_factor}. In the control region, both $\delta$ and $Re_{\theta}$ increase, i.e.\ the boundary layer is thickened by the control action. This is expected since the current controller imparts a positive net mass flow rate and it is known that uniform blowing increases the boundary layer thickness \citep{kametani2011direct, kametani2015effect, stroh2016global}. Note that $\delta$ starts to increase before the control slot is reached, and this is due to the upstream effect of the pressure gradient induced by the actuation, as will be discussed below. The shape factor $H$ is also increased above the control slot, in agreement with the findings from uniform blowing \citep{kametani2011direct, kametani2015effect}.

The effect of control action extends downstream region of the control slot. The shape factor $H$ gradually recovers and approaches the value of the uncontrolled flow. At the exit of the domain, $H \approx 1.5$ for both flows. At the moderate values of $Re_\theta$ shown in the middle figure (less than 1000), $H$ depends on the history of transition, \cite{Schlatter_Orlu_2012}. Although transition in the present work is due to interacting modes in the free-stream (and not due to tripping at the wall as in \cite{Schlatter_Orlu_2012}), the value at the exit is within the range of values reported by the aforementioned authors. Note also the constant negative shift of $Re_{\theta}$ plot for the controlled flow downstream of the slot; this is very similar to the shift reported in \cite{stroh2016global} for uniform blowing. 

\subsubsection{Pressure distribution}\label{subsec: pressure distribution}
\begin{figure}
\centering
\includegraphics[trim={0.2cm -0.cm -0.0cm -0.cm}, clip=true,width=0.8\textwidth]{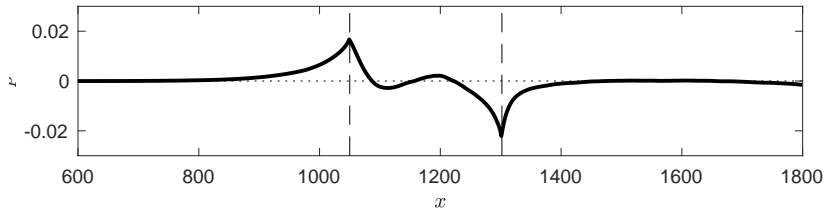}
\caption{Variation of mean wall pressure.}
\label{fig: pressure_contour}
\end{figure}

Figure \ref{fig: pressure_contour} shows the variation of the mean wall pressure. The mean pressure changes significantly near the entrance and the exit of the control slot. The time-average x-momentum equation at the wall can be written as,
\begin{equation}\label{eq: x-momentum equation at wall}
\left.\frac{\partial \overline{p}}{\partial x}\right\vert_w=\left.\frac{1}{Re}\frac{\partial ^2 \overline{u}}{\partial y^2}\right\vert_w-\left. \overline{v}_{w}\frac{\partial \overline{u}}{\partial y}\right\vert_w-\left.\overline{v_w^{\prime}\frac{\partial u^{\prime}}{\partial y}}\right\vert_w
\end{equation}
where the overbar denotes the average in time and the spanwise direction ($z$) and the prime ($^{\prime}$) denotes the fluctuation. Upstream and downstream of the control slot the pressure increases in the streamwise direction, i.e. there is an adverse pressure gradient (APG). This is because at these places $v_w=0$ and $\partial ^2 \overline{u} / \partial y^2 \vert_w >0$. The same is reported by \cite{park1999effects}, who applied uniform blowing to a turbulent boundary layer. The presence of the adverse pressure gradient upstream of the slot explains the upsteam effect of the control action mentioned earlier. Above the slot, $\overline{v}_{w}\left(\partial \overline{u}/\partial y\right)$ is always positive because $\overline{v}_w>0$ as shown in figure \ref{fig: spanwise_averaged_vw_period}. Our computations reveal that the third term in the right hand side of equation (\ref{eq: x-momentum equation at wall}) is much smaller than the second term, and therefore can be neglected. $\partial ^2 \overline{u} / \partial y^2 \vert_w $ is positive above the control slot and its magnitude is comparable to that of $-\overline{v}_{w}\left(\partial \overline{u}/\partial y\right)$. Therefore the sign of the pressure gradient depends on the relative size of these two terms. In the work of \cite{park1999effects}, the viscous term was much smaller as they used relatively large uniform blowing velocity, consequently they had favourable pressure gradient (FPG) over the entire control slot. In the present case, as seen in figure \ref{fig: pressure_contour}, the pressure gradient is initially favourable, and is followed by a weak adverse pressure gradient, starting from about $x=1100$, which is the location where $\overline{v}_w$ is decreasing to the local minimum (figure \ref{fig: spanwise_averaged_vw_period}). After $x=1200$, the pressure starts to decrease again as $\overline{v}_w$ keeps increasing. Near the exit of the domain, there is a small deviation of pressure from $0$,  probably due to the effect of the convective outlet boundary condition.

\subsubsection{Turbulent profiles}\label{subsec: turbulent profiles}

\begin{figure}\centering
\sidesubfloat[]{
\includegraphics[trim={-0.3cm -0.cm -0.10cm -0.cm}, clip=true,width=0.29\textwidth]{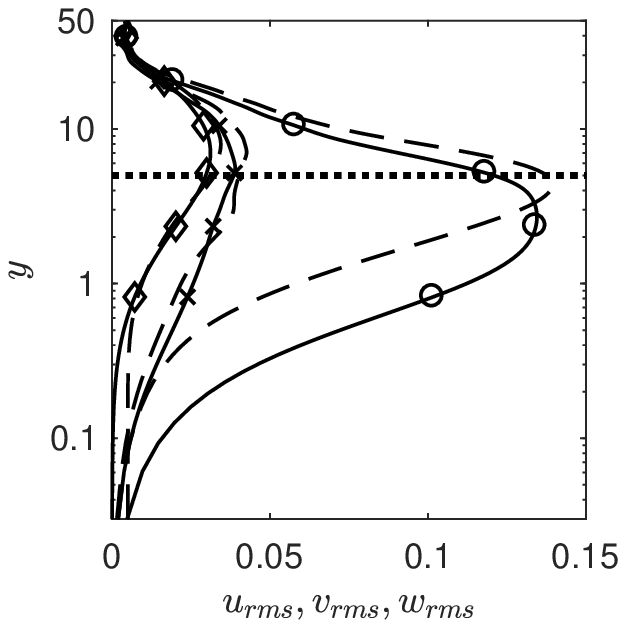}}
\sidesubfloat[]{
\includegraphics[trim={-0.3cm -0.cm -0.10cm -0.cm}, clip=true,width=0.29\textwidth]{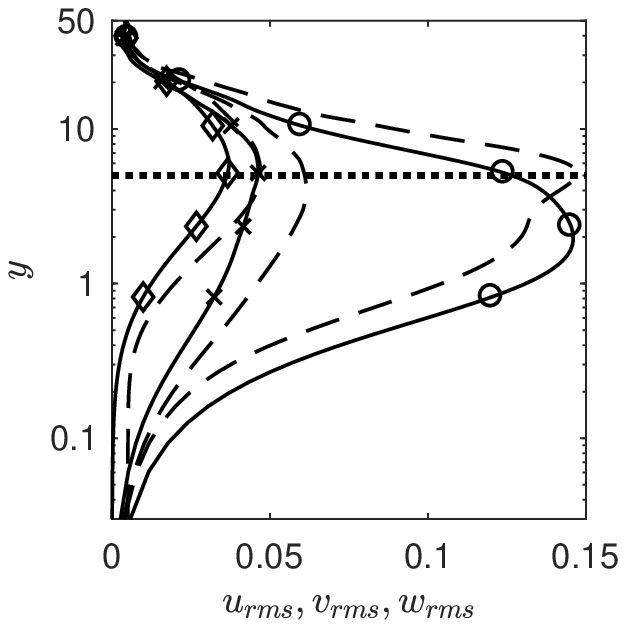}}\sidesubfloat[]{
\includegraphics[trim={-0.3cm -0.cm -0.10cm -0.cm}, clip=true,width=0.29\textwidth]{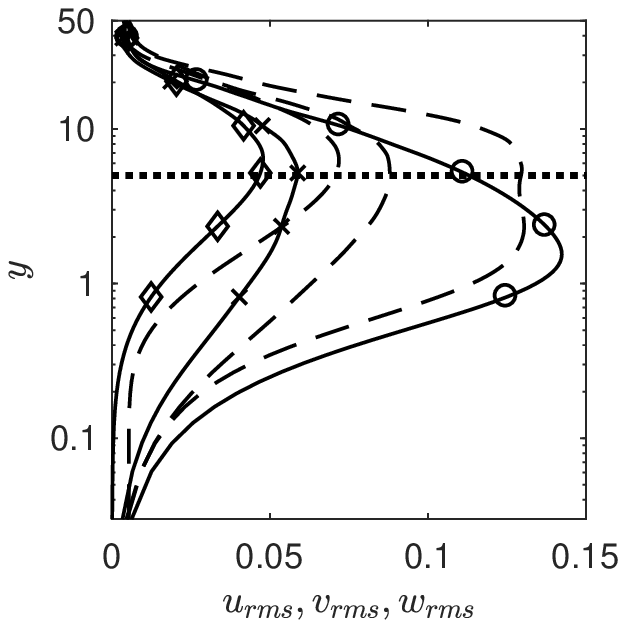}}\quad
\sidesubfloat[]{
\includegraphics[trim={0.cm -0.cm -0.10cm -0.cm}, clip=true,width=0.29\textwidth]{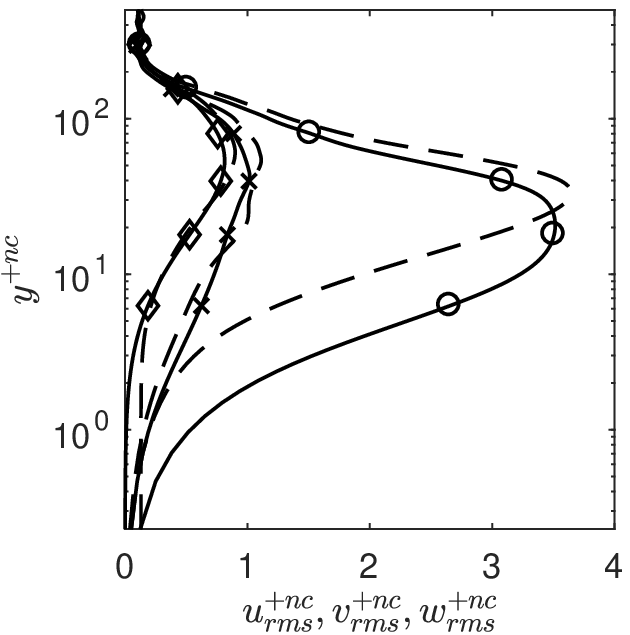}}
\sidesubfloat[]{
\includegraphics[trim={0.cm -0.cm -0.10cm -0.cm}, clip=true,width=0.29\textwidth]{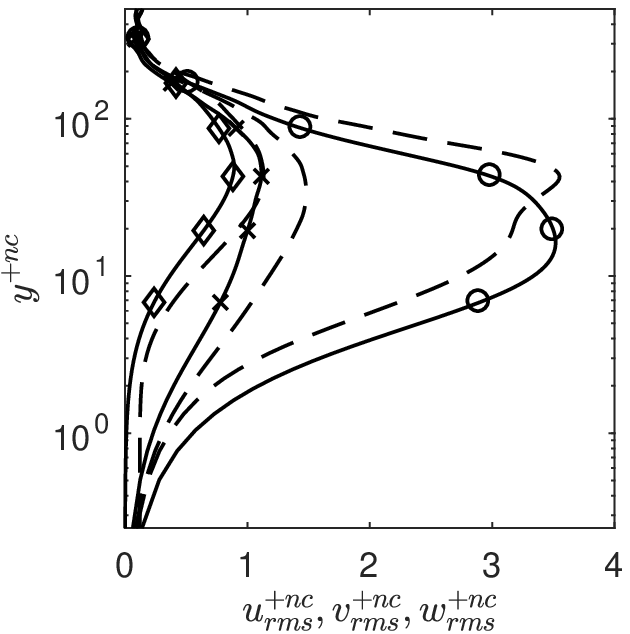}}\sidesubfloat[]{
\includegraphics[trim={0.cm -0.cm -0.10cm -0.cm}, clip=true,width=0.29\textwidth]{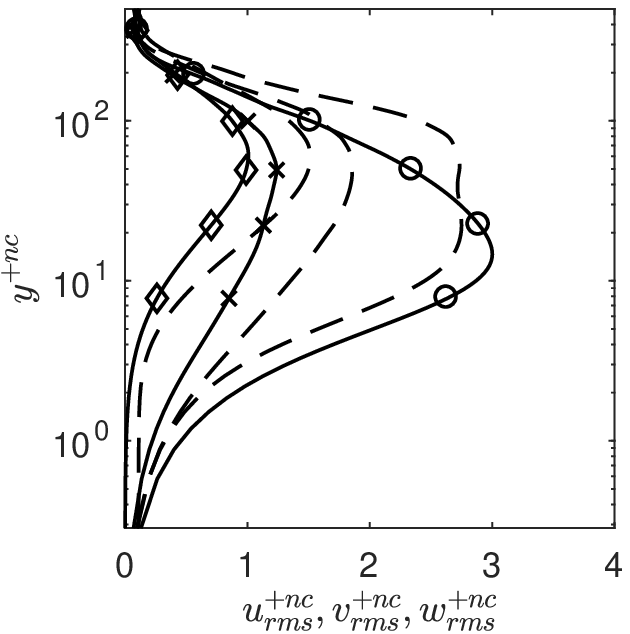}}\quad
\sidesubfloat[]{
\includegraphics[trim={0.cm -0.cm -0.10cm -0.cm}, clip=true,width=0.29\textwidth]{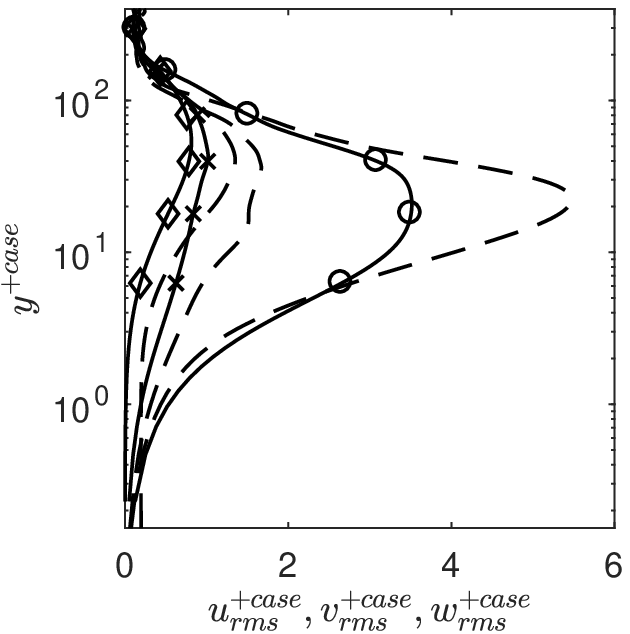}}
\sidesubfloat[]{
\includegraphics[trim={0.cm -0.cm -0.10cm -0.cm}, clip=true,width=0.29\textwidth]{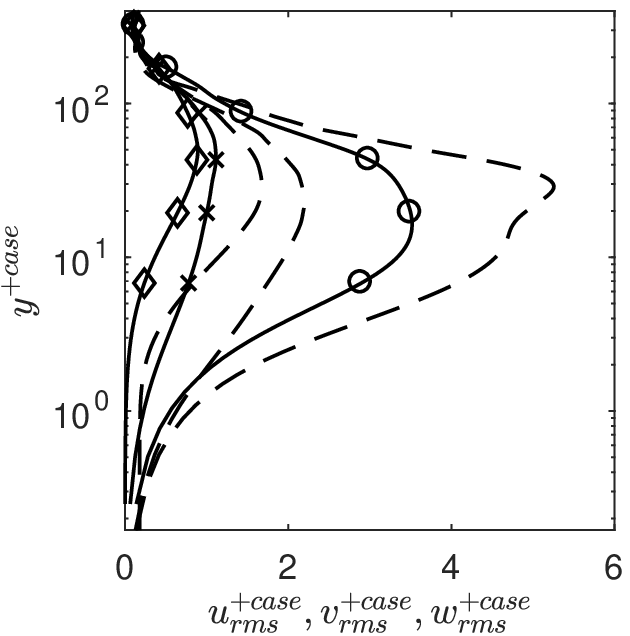}}\sidesubfloat[]{
\includegraphics[trim={0.cm -0.cm -0.10cm -0.cm}, clip=true,width=0.29\textwidth]{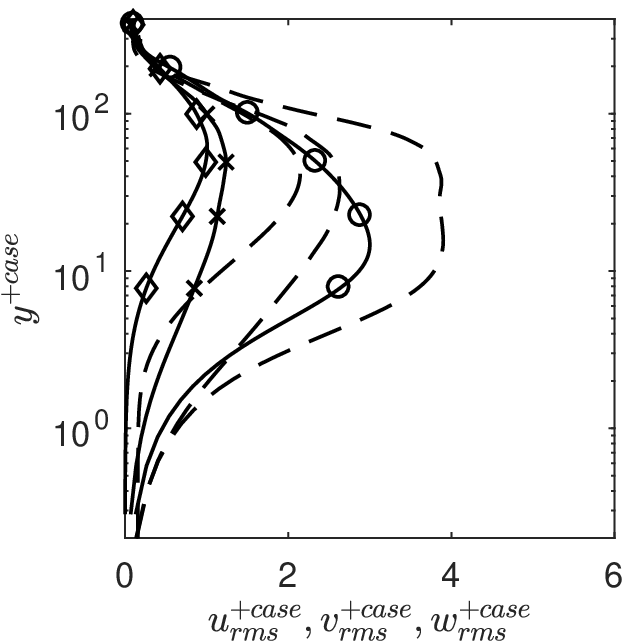}}
\caption{\label{fig: mean_rms_three_locations} Root mean square of the velocity fluctuations profile at three locations: first column, $x=1075$; second column, $x=1130$; third column, $x=1250$. First row, physical variables; second row, variables in wall units based on local uncontrolled friction velocity; third row, variables in wall unites based on case-specific, local friction velocity. $\circ$, $u_{rms}$, $\diamond$, $v_{rms}$, $\times$, $w_{rms}$. Solid line, uncontrolled flow; dashed line, controlled flow. The dotted line in the top row represents the extent of the region in which the cost function is defined.}
\end{figure}

In this section, the control effect on second order turbulent statistics is examined. Figure \ref{fig: mean_rms_three_locations} shows the rms profiles of the three velocity fluctuations at the same streamwise locations. The variables take physical values (first row), or are expressed in wall units based on local uncontrolled (second row) or case-specific, local friction velocity (third row).

In the uncontrolled flow, the wall-normal location for maximum $u_{rms}$ gradually reduces from $y=1.7$ ($0.135\delta$) at $x=1075$ to $y=1.54$ ($0.07\delta$). This is due to the fact that the transition process is initiated at the edge of the boundary layer and progressively penetrates inside the boundary layer, as already mentioned. In the controlled case, the values of $u_{rms}$ are reduced close to the wall (within the region where the cost function is defined) at all three locations; the reduction is largest at $x=1075$. Further away from the wall however, the values are increased, and more so in the two downstream locations.  When normalised by the case-specific (i.e. actual) local friction velocity, $u_{rms}^{+case}$ is increased all along the wall-normal direction, expect near the wall at $x=1075$. The enhanced $u_{rms}^{+case}$ is due to the fact that the friction velocity is decreased, as will be demonstrated later. 

The wall-normal turbulence intensity is positive at the wall due to the unsteady actuation velocity. At the first location $x=1075$, $w_{rms}$ is slightly decreased close to the wall while away from the wall slightly increased. $v_{rms}$ is only increased at peak. At two downstream locations, $x=1130$ and $x=1250$, the behaviour of $v_{rms}$ and $w_{rms}$ is different compared to the upstream point. All values increase in the wall-normal direction, which is also found in uniform blowing \citep{sumitani1995direct}. The overall effect is that in the right half of the slot turbulence intensity increases, and more so as the downstream end of the slot is approached. 

\begin{figure}\centering
\sidesubfloat[]{
\includegraphics[trim={0.cm -0.cm -0.20cm -0.cm}, clip=true,width=0.29\textwidth]{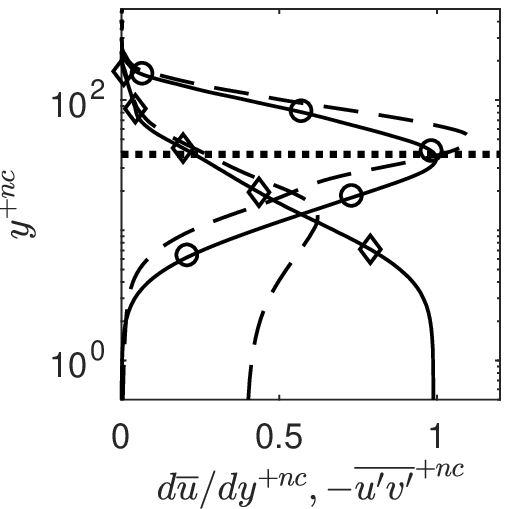}}
\sidesubfloat[]{
\includegraphics[trim={0.cm -0.cm -0.20cm -0.cm}, clip=true,width=0.29\textwidth]{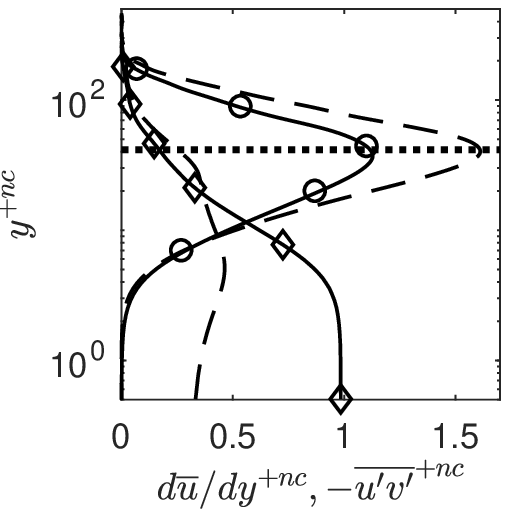}}\sidesubfloat[]{
\includegraphics[trim={0.cm -0.cm -0.20cm -0.cm}, clip=true,width=0.29\textwidth]{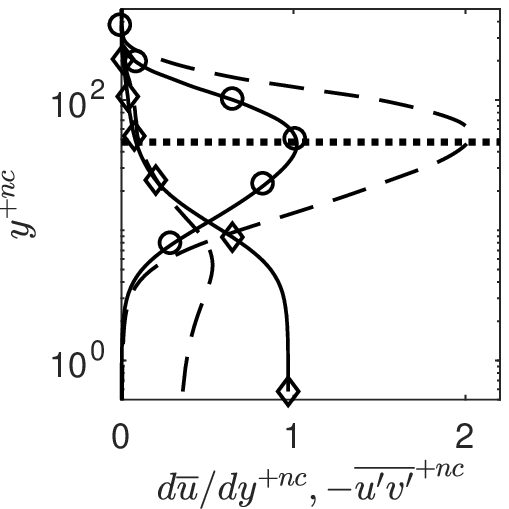}}
\caption{\label{fig: mean_stress_three_location} Reynolds stress $-\overline{u'v'}^{+nc}$ ($\circ$) and viscous shear stress $d\overline{u}/dy^{+nc}$ ($\diamond$) profiles at three locations: (a) $x=1075$; (b) $x=1130$; (c) $x=1250$. Solid line, uncontrolled flow; dashed line, controlled flow. All quantities non-dimensionalized by the local uncontrolled friction velocity.}
\end{figure}

Differences are also found in the profiles of the Reynolds and viscous shear stresses, shown in figure \ref{fig: mean_stress_three_location}. To facilitate comparison, the variables are expressed in wall units based on the local friction velocity of the uncontrolled flow. At all three locations, the viscous part $d\overline{u}/dy^{+nc}$ is decreased close to the wall, which is a direct result of the reduction of the friction velocity. The behaviour of Reynolds stress is different at the first location compared to the other two locations. At $x=1075$, $-\overline{u'v'}^{+nc}$ is reduced close to the wall. At $x=1130$, the controlled $-\overline{u'v'}^{+nc}$ is very close to that of the uncontrolled flow in the near wall region, while further downstream at $x=1250$, it is increased in the controlled case. Away from the wall, $-\overline{u'v'}^{+nc}$ is increased in all three locations. The behaviour of $-\overline{u'v'}^{+nc}$ at $x=1075$, i.e. decreased close to the wall and increased away from the wall, is also observed when using a blowing-only opposition control by \cite{pamies2007response}. On the other hand $-\overline{u'v'}^{+nc}$ at $x=1130$ and $x=1250$ is similar to those obtained from uniform blowing \citep{kametani2011direct, pamies2007response}. 

\subsection{Control effect downstream of the slot}\label{sec: control effect downstream of the slot}
\begin{figure}[!htbp]\centering
\includegraphics[trim={0cm 0cm 0cm 0cm},clip=true,width=0.95\textwidth]{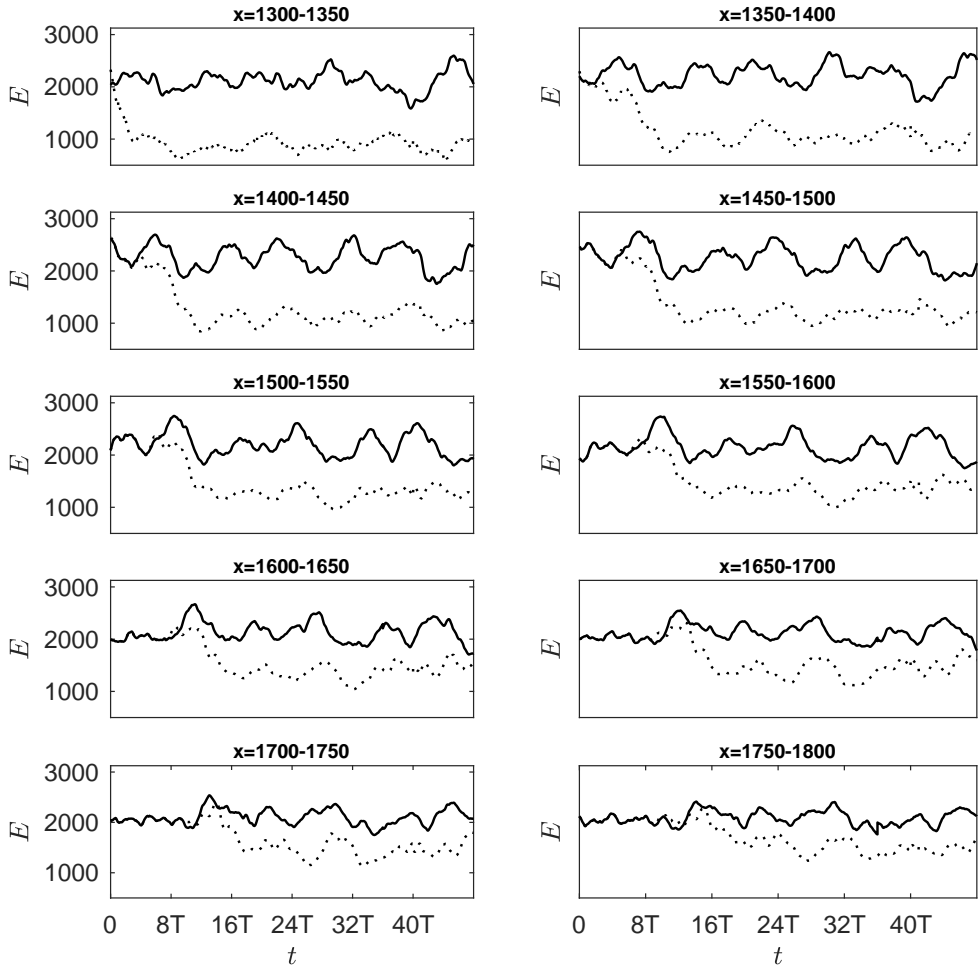}
\caption{\label{fig: Energy_flow2} Evolution of energy  downstream of the control slot. Each subplot corresponds to a segment of length $\Delta x=50$; the start and end positions are indicated at the top. Solid line, uncontrolled case; dashed line, controlled case.} 
\end{figure}

To investigate how the control effect propagates, we divide the distance between the downstream end of the control slot until the end of the computational domain in small segments of length $\Delta x=50$. The variation of $E$ with time is computed in each segment (in the wall normal and spanwise directions the segment dimensions are $0<y<5$, $0<z<90$ respectively). The results are displayed in figure \ref{fig: Energy_flow2}. $E$ is reduced immediately at the first segment $x=1300-1350$ and as the control effect propagates, reduction also appears downstream at later time instants. This shows that the control effect does affect the flow downstream of actuation slot. In fact, the flow energy is reduced right until the end of the domain. 

\begin{figure}\centering
\sidesubfloat[]{
\includegraphics[trim={0.cm -0.cm -0.20cm -0.cm}, clip=true,width=0.45\textwidth]{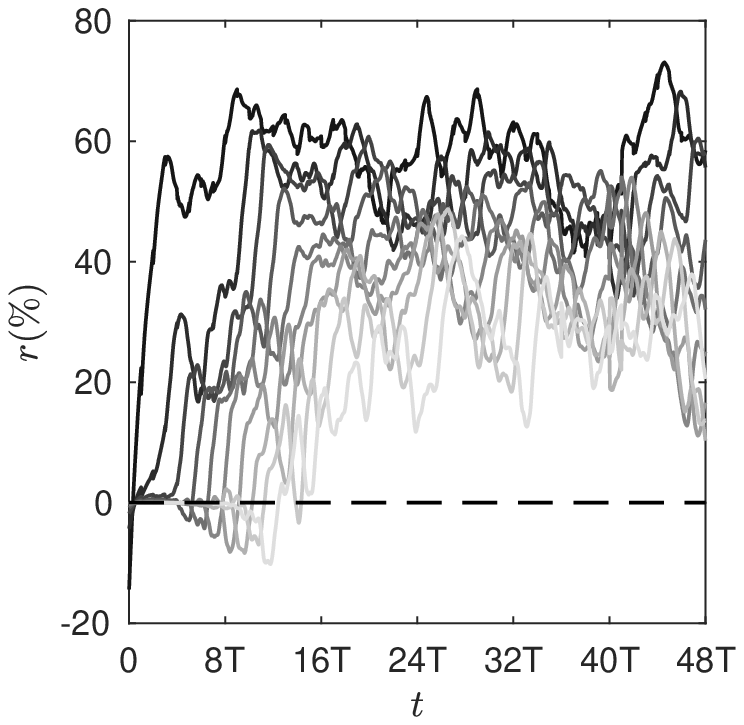}}
\sidesubfloat[]{
\includegraphics[trim={0.cm -0.cm -0.20cm -0.cm}, clip=true,width=0.45\textwidth]{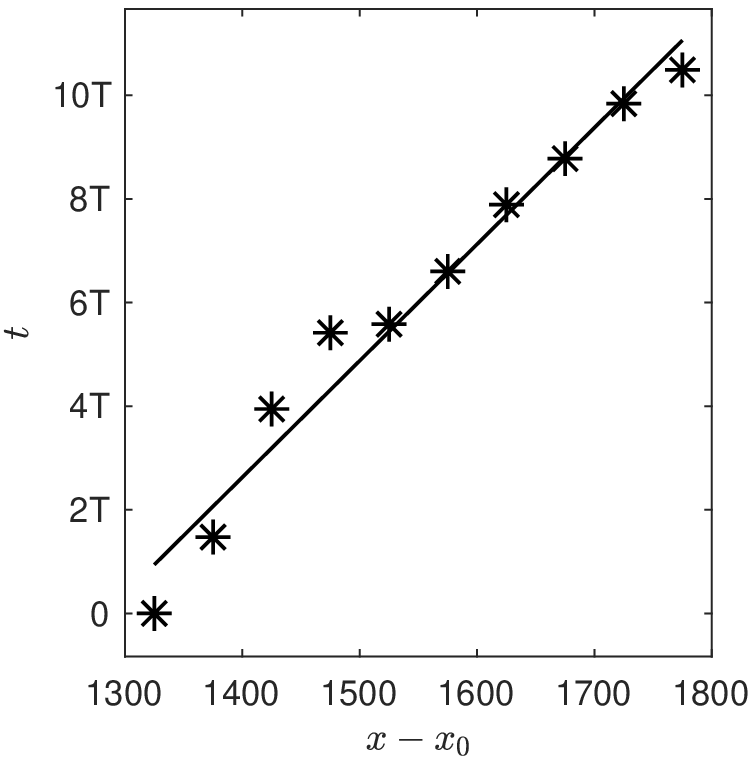}}
\caption{\label{fig: Energy_flow_reduction} (a) energy reduction rate $r=\left( E_{nc}-E_{c}\right)/ E_{nc}$ as a function of time; line color from dark to light corresponds to increasing starting $x$ of the 10 segments of figure \ref{fig: Energy_flow2}. (b) space-time diagram of the point when $\vert r \vert \geqslant 3 \%$ (indicated by a dash horizontal line in (a)).}
\end{figure}

The propagation of the control effect can be seen in figure \ref{fig: Energy_flow_reduction}, where the evolution of the energy reduction rate for each segment, defined as $r(t)=\left( E_{nc}-E_{c}\right)/ E_{nc}$, is plotted. The reduction rate varies from an average of $60\%$ in the first segment $x=1300-1350$ (closest to the exit of control slot) to an average of about $30\%$ near the end of the domain (furthest from the control slot). On the right a space-time diagram of the point at which $\vert r \vert \geqslant 3 \%$ is shown. The linear trend suggests that the actuation effect propagates at an average speed of 0.74$U_{\infty}$, which is very close to the convection speed of the streaks upstream as mentioned before.

\begin{figure}[!htbp]\centering
\includegraphics[trim={0cm 0cm 0cm 0cm},clip=true,width=0.7\textwidth]{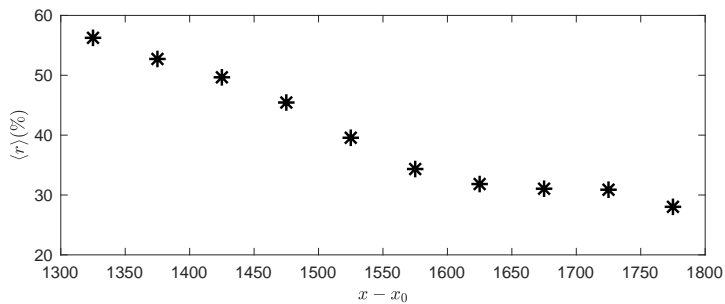}
\caption{\label{fig: Mean_energy_reduction} Mean energy reduction rate (averaged from $32T$ to $48T$) against distance $x$ downstream of the control slot.}
\end{figure}

After the control effect has reached the end of the domain and the flow has stabilised, the time-average (between $32T$ and $48T$) energy reduction rate downstream of the control slot can be computed. The results are shown in figure \ref{fig: Mean_energy_reduction}, and as expected, the energy reduction rate decreases further downstream. The computational domain is not big enough however to capture the full recovery to the uncontrolled state.
 
\begin{figure}[!htbp]\centering
\includegraphics[trim={0cm 0cm 0cm 0cm},clip=true,width=0.95\textwidth]{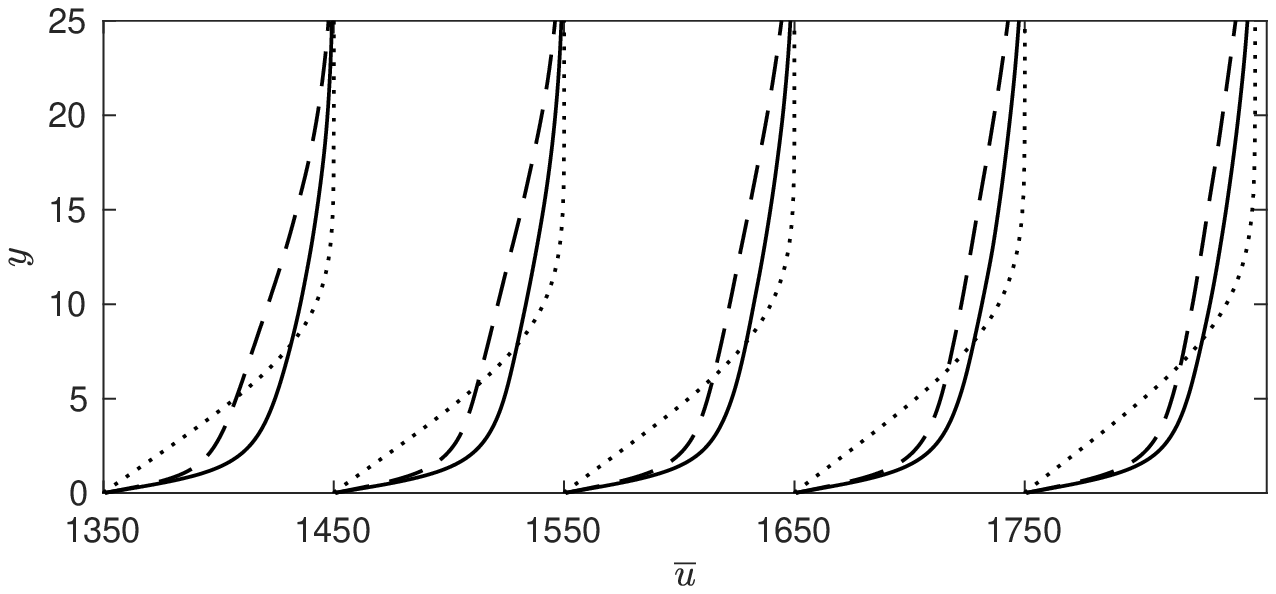}
\caption{\label{fig: mean_v_downstream} Mean velocity profile. Solid line, uncontrolled flow; dashed line, controlled flow; dotted line, Blasius solution. Streamwise axis shows the downstream location of each profile.}
\end{figure}

Attention is turned to the velocity profiles. Figure \ref{fig: mean_v_downstream} presents the mean velocity profiles at several downstream locations. The change in $\overline{u}$ in the near wall region ($y<2$) is small compared to the velocity profile above the control slots in figure \ref{fig: mean_v_three_locations}. Above $y=2$, there is large deviation of the controlled from the uncontrolled flow, but the difference becomes smaller further downstream. This is because the cost function is defined up to $x=1350$. There is a small difference of the profile slope at the wall, so we expect a small reduction on drag. The controlled $\overline{u}$ is reduced above the control slot (figure \ref{fig: mean_v_three_locations}) and downstream of the control slot (figure \ref{fig: mean_v_downstream}). The positive actuation introduces additional mass into the system and it is found that the controlled $\overline{v}$ is increased above the control slot to satisfy continuity.

\subsection{Effect on skin friction}
\begin{figure}\centering
\sidesubfloat[]{
\includegraphics[trim={-0.3cm -0.cm -0.20cm -0.cm}, clip=true,width=0.85\textwidth]{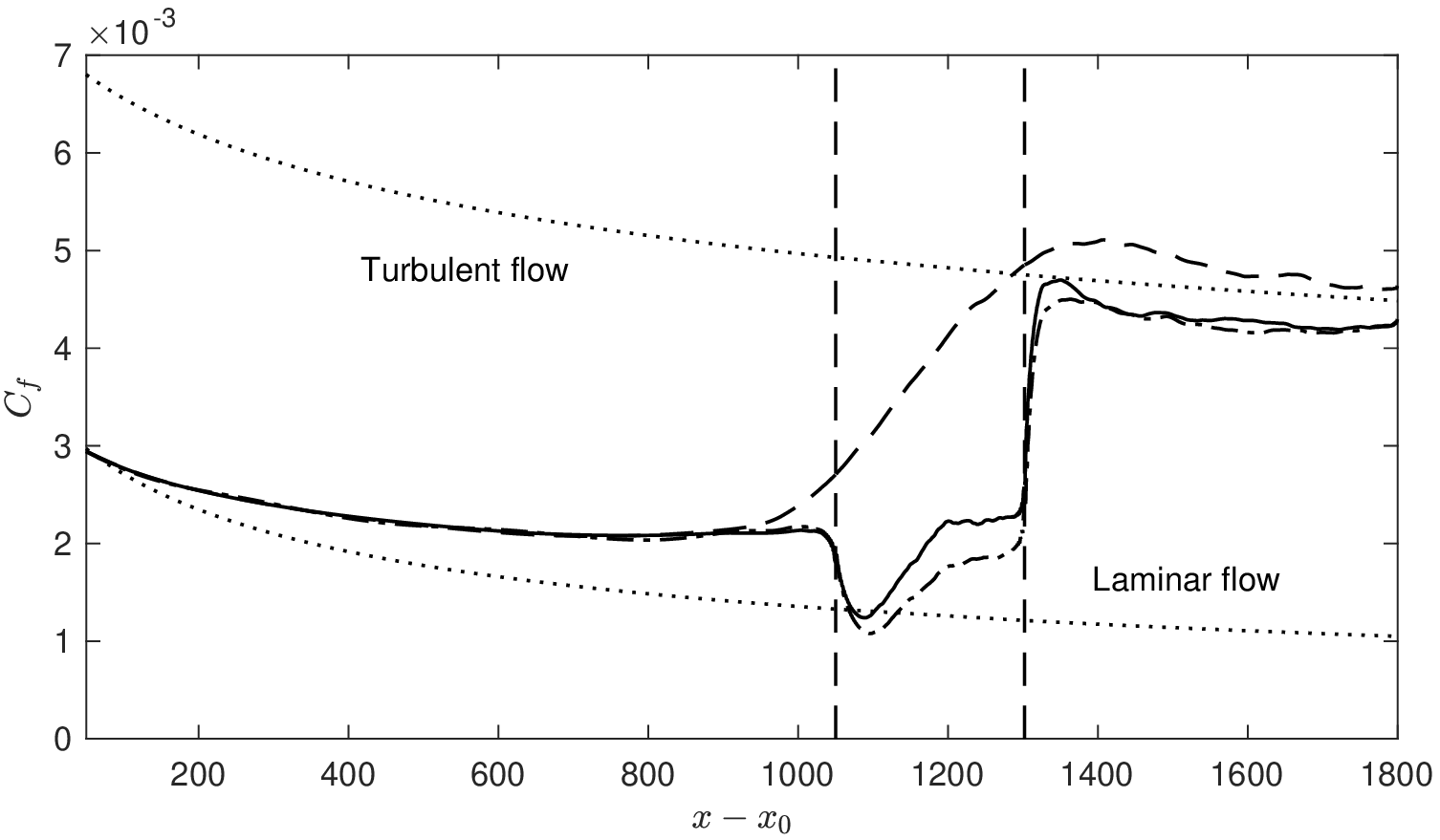}}\quad
\sidesubfloat[]{
\includegraphics[trim={0.cm -0.cm -0.20cm -0.cm}, clip=true,width=0.85\textwidth]{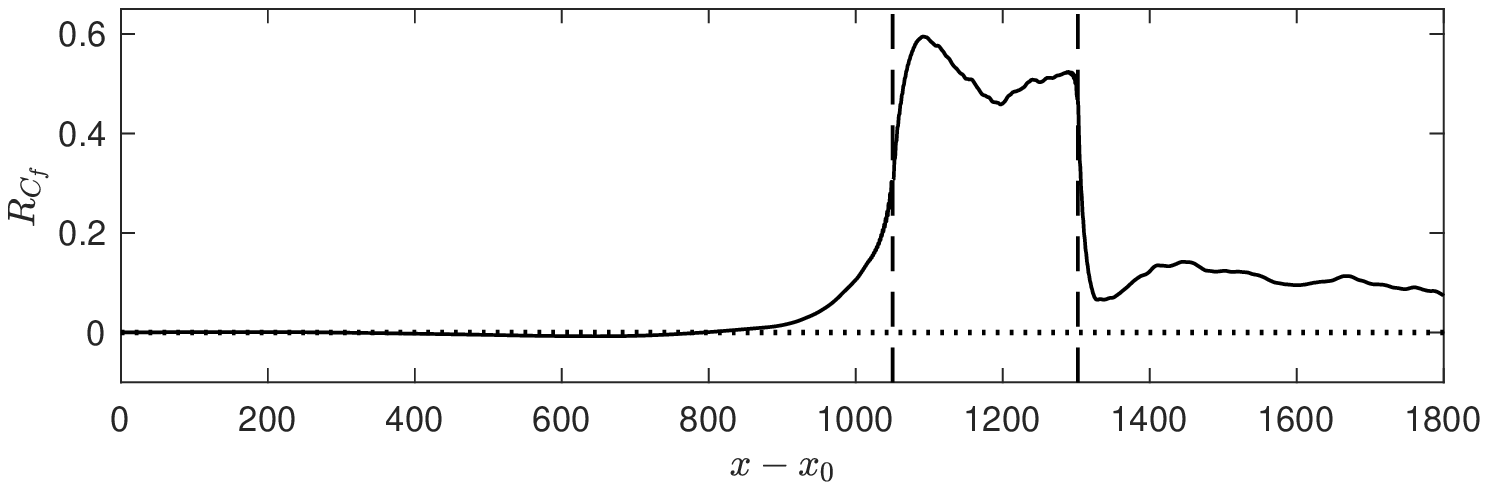}}
\caption{\label{fig: cf_mean_control_uncontrol} Mean skin friction profiles (a) and corresponding local drag reduction rate $R_{C_f}=(C_{f,nc}-C_{f,c})/C_{f,nc}$ (b). Solid line, \textcolor{black}{optimal controlled flow}; dashed line, \textcolor{black}{ 
uncontrolled flow}; dash-dot line, controlled flow using repeated $v_w$.}
\end{figure}

The time- and spanwise-averaged skin friction profile is shown in figure \ref{fig: cf_mean_control_uncontrol}. It can be seen that the controlled skin friction starts to deviate from the uncontrolled line upstream of the control region, which is also observed in other controlled flows \citep{park1999effects, pamies2007response, kim2006effects, kametani2015effect}. This is also in agreement with the mean pressure field shown in figure \ref{fig: pressure_contour}, where $\bar{p}$ at wall starts to increase from around $x=900$. The controlled $C_{f,c}$ reduces quickly inside the control slot and reaches the laminar level at around $x=1130$. Between $x=1130$ to $1200$, $C_{f,c}$ increases along the streamwise direction, at a rate similar rate to that of the uncontrolled flow. After $x=1200$, the controlled $C_{f,c}$ is almost constant until the end of the control region. At this location the actuation velocity $v_w$ increases linearly to counteract the transition. The local drag reduction rate $R_{C_f}=(C_{f,nc}-C_{f,c})/C_{f,nc}$ is also shown at the bottom subplot. Inside the control region, $R_{C_f}$ is between $45\%$ to $60\%$. 

Downstream of the control slot, $C_{f,c}$ recovers very quickly, but remains at a lower value than that of the uncontrolled flow up to the end of the domain. In terms of the average boundary layer thickness $\delta_c$ of the uncontrolled flow, the distance between the end of the control slot and the exit of the computational domain is  about $25\delta_c$. The average drag reduction rate in this region is $10\%$, with a maximum $R_{C_f}$ of $14.5\%$ at $7\delta_c$ and a minimum $R_{C_f}$ of $7.5\%$ at the exit. 

These results have similarities to those of \cite{stroh2016global}. The effect of uniform blowing was also extended significantly downstream of the control slot until the exit of the domain, as in our case. This is interesting because in our case turbulence is sustained by instabilities that are initiated inside the boundary layer and not close to the wall, as in \cite{stroh2016global}.  

It has been shown that the optimal control velocity is repeatable in a time scale equivalent to the period of the slow mode (figure \ref{fig: spanwise_averaged_vw_period}). The $C_{f,c}$ from optimal control shown in figure \ref{fig: cf_mean_control_uncontrol} (black dashed line) was averaged from $t=16T$ to $48T$, which is limited by the heavy computational cost for this control method. Although our results are in agreement with those of \cite{stroh2016global} as mentioned above, we performed an additional check to confirm that the drag reduction observed downstream of the slot is indeed robust, and it is not an artifact of the short time-averaging window. To this end, the temporal and spatial evolution of the optimal control velocity was stored, and then applied repeatedly for $64T$ (equivalent to 4 periods of the slow mode and two flow through times). 

\begin{figure}\centering
\includegraphics[trim={0cm 0cm 0cm 0cm},clip=true,width=0.9\textwidth]{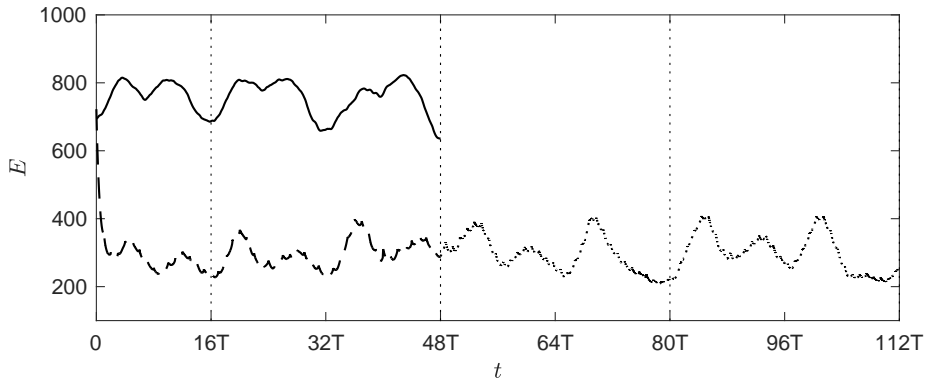}
\caption{\label{fig: energy_repeat}  Evolution of the flow energy defined in the control box. Black solid line, uncontrolled flow; thick dashed line, optimal controlled flow; thin dashed line, controlled flow using repeated application of $v_w$.}
\end{figure}

The evolution of flow energy in the same control box is shown in figure \ref{fig: energy_repeat}. The vertical dotted lines indicate repeated application of $v_w$, i.e. the optimal control velocity from $16T$ to $48T$ was repeated from $48T$ to $80T$ and $80T$ to $112T$. The figure demonstrates that the repeated application of $v_w$ is able to maintain the flow energy at a level very similar to that obtained from piece-wise optimisation. The controlled $E$ also has a clearly periodic behaviour over the time, as expected. 

The long time-averaged $C_{f,lc}$ from $16T$ to $112T$ (i.e. including both piece-wise optimal control and control using repeated $v_w$ profiles) is shown in figure \ref{fig: cf_mean_control_uncontrol} by the dash-dot line. It is somewhat surprising that in the control slot,  $C_{f,lc}$ is slightly lower than the $C_{f,c}$ from the receding horizon control. It must be borne in mind however that the objective of the optimal control is to minimise the flow energy, and not $C_f$. Most importantly, the results demonstrate that in the downstream region, the long time-averaged $C_{f,lc}$ remains reduced and has a very similar value as the $C_{f,c}$ from receding horizon control.

\subsection{Correlation between actuation and the flow field}\label{sec: Correlation between actuation and the flow field}

In the present work, the control velocity distribution was obtained solely based on the objective function and the governing equations. In this section we explore the relationship between the optimal actuation and the flow above the slot. 

\begin{figure}\centering
\sidesubfloat[]{
\includegraphics[trim={-0.cm -0.cm -0.20cm -0.cm}, clip=true,width=0.8\textwidth]{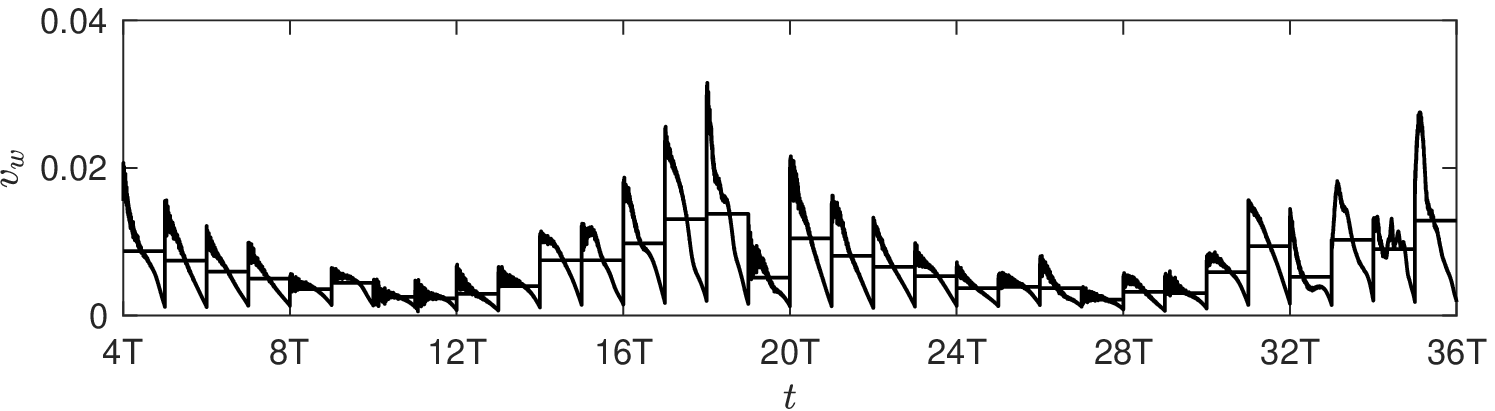}}\quad
\sidesubfloat[]{
\includegraphics[trim={0.cm -0.cm -0.20cm -0.cm}, clip=true,width=0.8\textwidth]{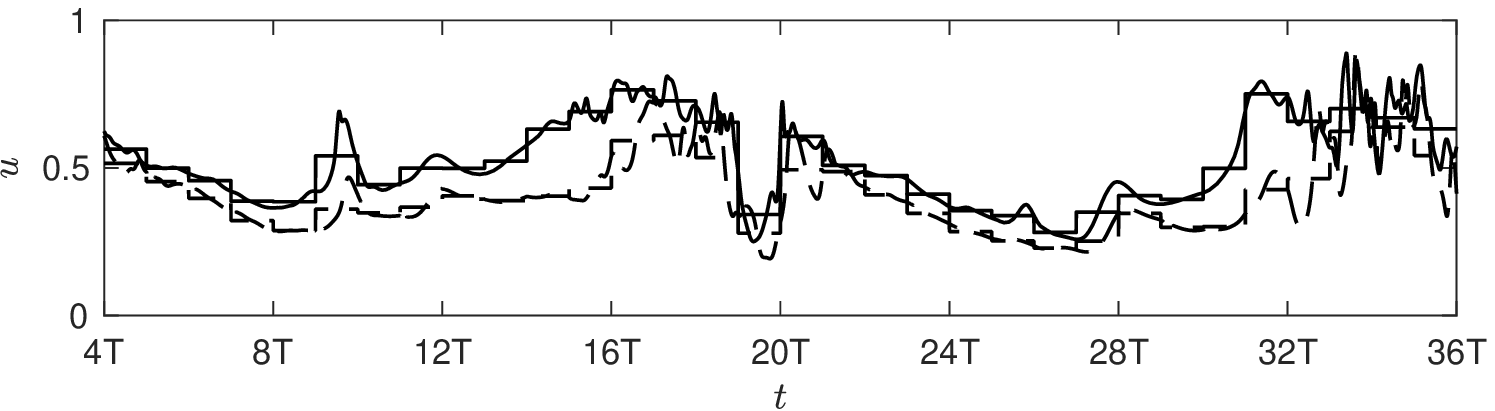}}
\caption{\label{fig: oridinal_data_x1080}  Time history of control velocity $v_w(x=1080, z=30)$ (a) and instantaneous streamwise velocity $u(x=1080, y=2.5, z=30)$ (b). Solid line, uncontrolled $u$; dashed line, controlled $u$; thin horizontal solid line, time averaged value over each optimisation interval.}
\end{figure}

\textcolor{black}{Figures \ref{fig: oridinal_data_x1080} shows the time history of $v_w$ and $u$ at $x=1080$, which is close to the inlet of the control slot. Velocity $u$ is recorded at a distance $y=2.5$ from the wall. The time is from $4T-36T$, which covers two complete periods of the slow inlet mode. It was shown earlier that at this location the flow contains distorted streaks. The control velocity $v_w$ and both uncontrolled and controlled $u$ exhibit periodic behaviour as a function of time and it is clearly that they are correlated.}

In order to explore quantitatively the relationship between the optimal actuation and the flow above the slot over time, we compute the two-point correlation coefficient between $v_w(x,z)$ and a general variable of interest $\phi(x,y,z)$; in the present case $\phi(x,y,z)$ represents velocities $u$ or $v$. The coefficient is defined as,
\begin{equation}
R_{v_w,\phi}^{t}(x,z;y)=\frac{\langle\left(v_w-\langle v_w \rangle_{t} \right)\left(\phi-v_w-\langle \phi \rangle_{t}\right)\rangle_t}{\left[\langle\left(v_w-\langle v_w \rangle_{t} \right)^2\rangle_t \langle\left(\phi-\langle \phi \rangle_{t} \right)^2\rangle_t\right]^{1/2}}
\end{equation}
The superscript $t$ denotes averaging in time from $4T$ to $36T$, and the correlation is computed using data from two periods of time, i.e. 32T. Note that due to the finite value of $T$, velocity $v_w$ decreases over time within each interval and rapidly increases at the beginning of next interval. On the other hand the velocity field is relatively smooth compared to $v_w$. The correlation coefficients are computed using data from each time instant from $4T$ to $36T$ and also from data averaged within each optimisation interval (shown by horizontal lines in figure \ref{fig: oridinal_data_x1080} ). 

\begin{figure}[!htbp]\centering
\includegraphics[trim={4cm 7.6cm 4.5cm 2.5cm},clip=true,width=0.8\textwidth]{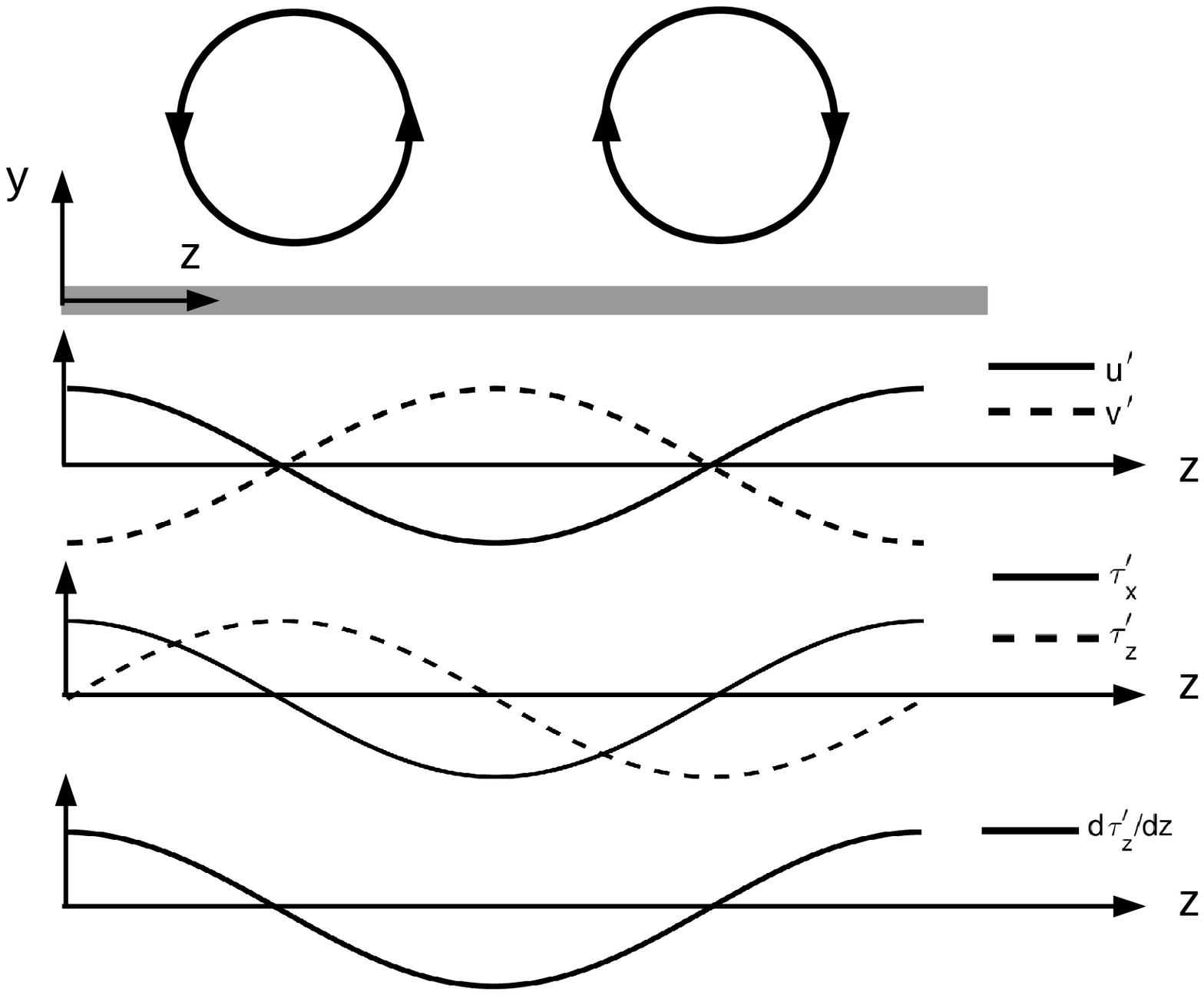}
\caption{Sketch of a pair of streamwise vortices and corresponding variation in velocity fluctuations ($u^{\prime},v^{\prime}$) at the height of vortex center and wall shear stresses $\tau_x^{\prime}, \tau_z^{\prime}$ \citep{naguib2010relationship}.}
\label{fig: vortex_pair_sketch}
\end{figure}

Before examining the correlation between the control velocity and flow field above it, we first recall in figure \ref{fig: vortex_pair_sketch} the flow properties corresponding to an idealized vortex pair. The largest (smallest) velocity fluctuations $u^{\prime}$ ($v^{\prime}$) at the height of the vortex center occurs in-between the vortices, while at the centre of the vortex they are zero. Note also that $u^{\prime}$ and $v^{\prime}$ are negatively correlated (leading to ejections and sweeps events). The spanwise variation of the streamwise shear stress $\tau_x$ is consistent with $u^{\prime}$. On the other hand, the largest $\tau_z$ is directly beneath the vortex center and therefore is offset from the peak $u^{\prime}$ position by a quarter of the average streak spacing \citep{naguib2010relationship}. 

\begin{figure}\centering
\sidesubfloat[]{
\includegraphics[trim={-0.cm -0.cm 0.20cm -0.cm}, clip=true,width=0.45\textwidth]{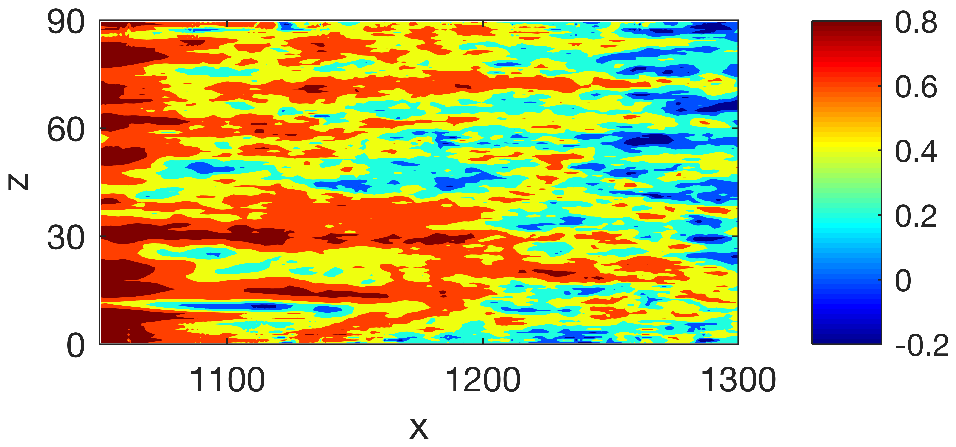}}
\sidesubfloat[]{
\includegraphics[trim={-0.cm -0.cm 0.20cm -0.cm}, clip=true,width=0.45\textwidth]{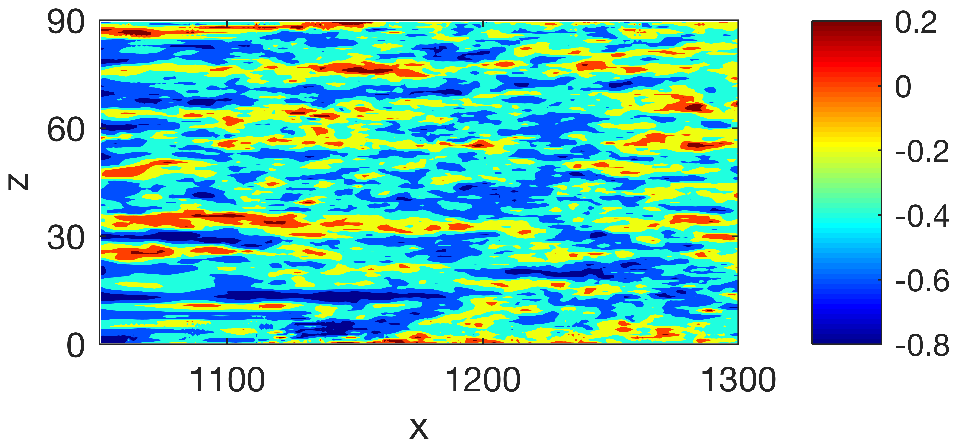}}
\caption{\textcolor{black}{\label{fig: correlation_contour_above_u1} Contours of correlation between actuation velocity $v_w(x, z)$ and averaged data for (a) instantaneous streamwise velocity $u(x, y=2.5, z)$, (b) instantaneous wall normal velocity $v(x, y=4, z)$. }}.
\end{figure}

\textcolor{black}{Figure \ref{fig: correlation_contour_above_u1} shows contours of $R_{v_w, \phi}^{t}$ over the slot for $\phi=u$ and $\phi=v$ extracted at $y=2.5$. The correlation was computed using averaged data from $4T$ to $36T$. When averaged data are used, the sample is much smaller (there is one sample per interval, corresponding to the horizontal line of figure \ref{fig: oridinal_data_x1080}). It can be seen that for $\phi=u$, there is strong positive correlation over the entire slot; it is only towards the downstream end where some patches with negative correlation appear. $R_{v_w, u}^t$ has a strong spatial dependence that weakens along the streamwise direction while the correlation of $R_{v_w, v}^t$ remains more uniform over the entire slot. The values of $R_{v_w, v}^t$ are negative in most part of control region, indicative of opposition control.}

\begin{figure}
\centering
\includegraphics[trim={-0.cm -0.cm 0.0cm -0.cm}, clip=true,width=0.8\textwidth]{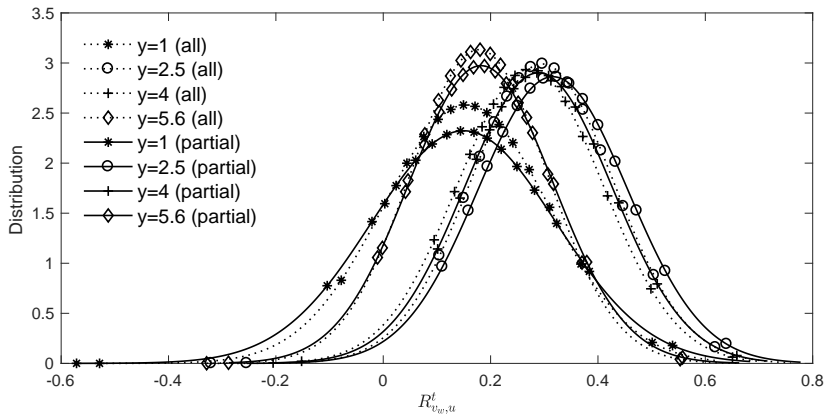}
\caption{\label{fig: correlation_pdf_above_u1} Distribution of the correlation coefficient $R_{v_w, u}^{t}$ between $v_w(x, z)$ and $u(x, y, z)$ at different wall-normal locations.}
\end{figure}

In order to remove the effect of the transient behavior at the beginning of each interval due to the instability of the adjoint equations, we recomputed the correlations using the non-averaged data but discarding the first $150$ time steps (equivalent to $t=12$) in each optimisation interval. We further processed the correlation over the entire slot and produce a (smoothed-out) normalised histogram of the correlation distribution. Figure \ref{fig: correlation_pdf_above_u1} shows the histogram of $R_{v_w, u}^t$ for non-averaged data (denoted by 'all') and non-averaged data with first 150 time steps removed (denoted by 'partial') for various wall-normal locations. The results show that the transient in $v_w$ has small effect on the correlation distribution. There is positive correlation between $v_w$ and $u$ at all four locations and $R_{v_w, u}^t$ is larger at $y=2.5$ and $y=4$ than at $y=1$ and $y=5.6$. This wall-normal dependence is very similar to $R_{v_w, u}^s$ (figure \ref{fig: correlation_space_y}). The histogram peaks at average $R_{v_w, u}^s\approx 0.3-0.35$.

\begin{figure}
\centering
\includegraphics[trim={-0.cm -0.cm 0.0cm -0.cm}, clip=true,width=0.8\textwidth]{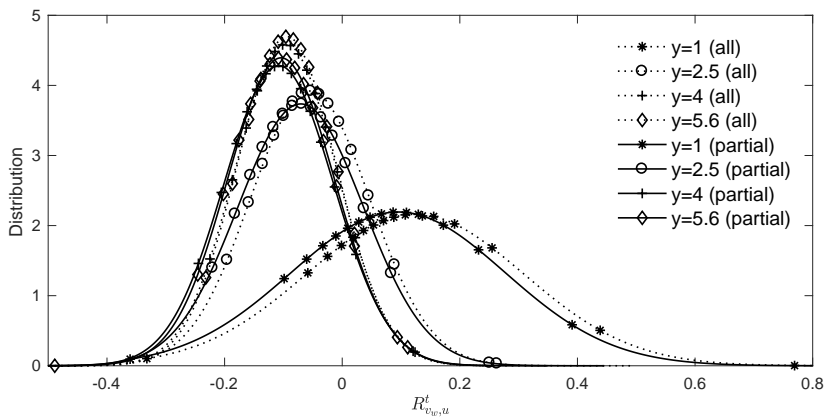}
\caption{Distribution of the correlation coefficient $R_{v_w, v}^{t}$ between $v_w(x, z)$ and $v(x, y, z)$ at different wall-normal locations.}
\label{fig: correlation_pdf_above_u2}
\end{figure}

Figure \ref{fig: correlation_pdf_above_u2} shows the distribution of $R_{v_w, v}^{t}$ at the same four wall normal locations. The correlation is negative at 3 locations, apart from $y=1$ where it is slightly positive. This is because this location is close to the wall and $v$ is strongly affected by the actuation. As for $R_{v_w, u}^{t}$, removing the transient data results in slight changes. The histogram peaks at average $R_{v_w, v}^s\approx -0.10$.

The standard opposition control \citep{choi1994active} imposes the opposite of the wall-normal velocity at a distance away from the wall (detection plane) with the aim to counteract the up-and-down motion induced by the vortex. In this case the correlation at the detection plane is $R_{v_w, v}^s=-1$. The predicted negative sign of the $R_{v_w, v}^s$ correlation in our case indicates that opposition control is at play, but the small value of the correlation indicates that it significantly deviates from the standard opposition control. The correlation is positive with the streamwise velocity, so the optimal wall actuation is broadly consistent with the streamwise vortex model shown in figure \ref{fig: vortex_pair_sketch}.

\begin{figure}\centering
\includegraphics[trim={0cm 0cm 0cm 0cm},clip=true,width=0.6\textwidth]{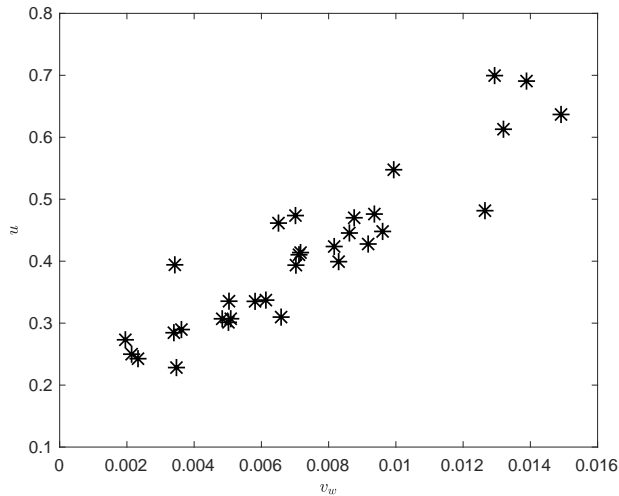}
\caption{\label{fig: joint_pdf_example} Scatter plot of time-average (over one interval) $v_w$ at $x=1071$, $z=63$ and $u$ at $x=1071$, $y=2.5$, $z=63$, at this location $R_{v_w, u}^{t}=0.92$.}
\end{figure}

Recall that the current actuation has mean positive velocity (figure \ref{fig: spanwise_averaged_vw_period}). Figure \ref{fig: joint_pdf_example} presents a scatter plot of $v_w$ and $u$ from averaged data at one particular location with averaged $R_{v_w, u}^{t}=0.92$ as an example. A positive $R_{v_w, u}^{t}$ implies that a larger positive $v_w$ is applied to area with positive $u^{\prime}$, while smaller positive $v_w$ is applied to area with negative $u^{\prime}$.  Therefore the actuation is closer to the blowing-only opposition control of \cite{pamies2007response}, where the suction part from the opposition control is removed. \cite{chang2002viscous} showed that the effectiveness of standard opposition control reduces as Reynolds number increases in channel flow. \cite{pamies2007response} demonstrated the blowing-only opposition control can improve drag reduction efficiency when compared to the classic opposition control as well as uniform blowing with same mean control velocity. Another similar example was the recent experimental work of \cite{abbassi2017skin}, in which wall-normal jet flow was injected in regions where high-speed streamwise velocity fluctuations were presented. The jet operated in the blowing-only mode.

In section \ref{sec: control effect inside the controller region} it was found that the turbulent profiles (r.m.s. of velocity fluctuations, Reynolds stress) have different behaviour over the control slot. In particular, the controlled flow at $x=1075$ is different from that at $x=1130$ and $x=1275$. In figure \ref{fig: correlation_contour_above_u1} is can be seen $R_{v_w, u}^{t}$ is larger in the upstream side of the control slot, especially upstream of $x=1100$. As discussed above, this indicates the control action is similar to blowing-only opposition control in this region. This is supported by the behaviour of the controlled turbulent profiles at $x=1075$ is very similar to those obtained by blowing-only opposition control \citep{pamies2007response}.  The blowing-only opposition control is designed to counteract the sweep events only.  \cite{wallace2016quadrant} reported that close to the wall, for $y^+<15$, the sweep events contribute considerably more to the total Reynolds stress than the ejection events. In figure \ref{fig: mean_stress_three_location}, at $x=1075$ $-\overline{u'v'}^{+nc}$ is largely reduced near the wall at $x=1075$. There is very small reduction in $-\overline{u'v'}^{+nc}$ at $x=1130$ while at $x=1250$ an increase is observed. This behavior agrees with the spatial distribution of $R_{v_w, u}^{t}$. 

In the rear part of the control slot when $R_{v_w, u}^{t}$ is weak, the mean positive actuation indicates the controller might work in a similar way as the uniform blowing. This is also proved by the behaviour of turbulent statistics in this region.

The change of the control mechanism over the slot is related to the flow activity in the control region. As shown in section \ref{sec: Flow energy}, near the inlet of the control region, there are mainly distorted streaks entering from upstream and the controller tends to counteract the motion of the streaks (vortices) using blowing-only opposition control. This is effective as the flow energy initially reduces along $x$ (figure \ref{fig: Energy_flow_spatial}). As mentioned, the transition still occurs in the controlled flow after $x=1100$ and there are turbulent structures, but with reduced strength. In this region ($x>1100$), the optimal results reveal that the best solution to control the flow is to impose strong positive actuation. Although the controlled flow properties are similar to those obtained from uniform blowing, the distribution of optimal $v_w$ in this region is not uniform. In \citet{xiao2017nonlinear} it was shown that a variable $v_w$ is more efficient in reducing the objective function than uniform blowing in a single interval. The same is also demonstrated in the receding horizon control.

\section{Conclusions}\label{sec:conclusions}
In the present work, a nonlinear optimal control strategy is applied in a receding horizon framework in order to suppress bypass transition in a zero-pressure-gradient flat-plate boundary layer. The transition process is triggered by \textcolor{black}{a pair of} free stream vortical perturbations, which consist of two continuous Orr-Sommerfeld and Squire modes. The optimal control problem is solved using the Lagrange multiplier technique. The objective is to find the optimal blowing and suction velocity that results in the minimum of the weighted sum of energy of velocity perturbations around the Blasius profile and the actuation energy. The control slot is located in the late transition region, where turbulent spots break down, grow and merge into turbulence. Using the receding horizon approach, the control is applied for longer time so that time-averaged statistics can be examined in order to gain more insight into the control action.

The results show that the controller is very effective in reducing the objective function. The uncontrolled flow energy increases \textcolor{black}{monotonically} in the streamwise direction as a result of the transition process taking place over the control slot. On the other hand, the controlled flow energy initially decreases along the  streamwise direction near the beginning of the control slot, then increases towards the rear, but with a reduced strength. This spatial dependence results from the competition between the control action and the transition activity. The optimal control velocity has a positive net mass flow rate and its spatial distribution is found to reflect the transition process.

The control performance is further investigated through time-averaged statistics. Over the control slot, the controller does its duty, and brings the mean velocity towards the Blasius profile. The control effect propagates downstream of the slot up, right up to the end of the computational domain. An average drag reduction of $55\%$ and $10\%$ is achieved over the control slot and in the downstream region, respectively.

The correlations between the optimal control velocity and various flow properties above the slot are also examined. It is found that the actuation velocity is positively correlated with instantaneous streamwise velocity, especially near the upstream half of the control slot. This implies that the controller works similar to the blowing-only opposition control near the beginning of the slot, while near the rear side of the control region, the controller works similar to uniform blowing. This is supported by the second-order statistics of the controlled flow (rms and Reynolds stress). 

The present nonlinear optimal control strategy has been shown to be effective in suppressing bypass transition. Without any constraint on the mass flow rate of the actuation, the optimal control law results in a positive net mass flow rate. A natural extension of this work would be to enforce zero mass flow rate constraint, which is easier to implement in practice. 

\setcounter{secnumdepth}{0}
\section{Acknowledgments}
The simulations were performed in the CX2 facility of Imperial College as well as Archer (to which access was provided through the UK Turbulence Consortium grant EP/L000261/1).

\bibliographystyle{jfm}
\bibliography{ref_revised,ref_gp_revised}

\end{document}